# The Concepts and Applications of Fractional Order Differential Calculus in Modelling of Viscoelastic Systems: A primer

Mohammad Amirian Matlob[1], Yousef Jamali[1,2*]

[1] Biomathematics Laboratory, Department of Applied Mathematics, Tarbiat Modares University, Iran
[2] Computational physical Sciences Research Laboratory, School of Nano-Science, Institute for Research in Fundamental Sciences (IPM), Tehran, Iran

**Viscoelasticity and related phenomena are of great importance in the study of mechanical properties of material especially, biological materials. Certain materials show some complex effects in mechanical tests, which cannot be described by standard linear equation (SLE) mostly owing to shape memory effect during deformation. Recently, researchers have been applying fractional calculus in order for probing viscoelasticity of such materials with a high precision. Fractional calculus is a powerful tool for modeling complex phenomenon. In this tutorial based paper, we try present clear descriptions of the fractional calculus, its techniques and its implementation. The intention is to keep the details to a minimum while still conveying a good idea of what and how can be done with this powerful tool. We try to expose the reader to the basic techniques that are used to solve the fractional equations analytically and/or numerically. More specifically, modeling the shape memory phenomena with this powerful tool are studied from different perspectives, as well as presented some physical interpretation in this case. Moreover, in order to show the relationship between fractional models and standard linear equations, a fractal system comprising spring and damper elements is considered, and the constitutive equation is approximated with a fractional element. Finally, after a brief literature review, two fractional models are utilized to investigate the viscoelasticity of the cell, and the comparison is made among them, experimental data, and previous models. Verification results indicate that not only does the fractional model match the experimental data well, but it also can be a good substitute for previously used models.**

## Author Summary

**Fractional Calculus is a new powerful tool which has been recently employed to model complex biological systems with non-linear behavior and long-term memory .In spite of its complicated mathematical background, fractional calculus came into being of some simple questions which were related to the derivation concept; such questions as while the first order derivative represents the slope of a function, what a half order derivative of a function reveal about it? Finding answers to such questions, scientists managed to open a new window of opportunity to mathematical and real world, which has arisen many new questions and intriguing results. For example, the fractional order derivative of a constant function, unlike the ordinary derivative, is not always zero. In this tutorial-based paper it is sought to answer the aforementioned questions and to construct a comprehensive picture of what fractional calculus is, and how it can be utilized for modelisation purpose. The focus of the research has been on viscoelastic materials. After an extensive literature review of the concepts and application of this potent tool, a novel application of this tool is developed for simulating a dynamic system in order to investigate the mechanical behavior of a cell.**

*Corresponding author. Tel: +98-21-8288-4762
E-mail addresses: y.jamali@ipm.ir (Y.Jamali), m.amirianmatlob@modares.ac.ir (M.Amirian Matlob).

# Contents

# Introduction

The concept of derivative is the main idea of calculus. It shows the sensitivity to change of a function i.e. the rate or slope of a quantity. The current definition of derivative was suggested by Newton in 1666. Newton with a physical viewpoint of derivative interpreted the instantaneous velocity [1]. The intuition of researchers from derivative and integral is based on their geometrical or physical meaning, e.g. the first and second order derivative of displacement is called velocity and acceleration respectively (also jerk and jounce for 3th and 4th derivatives). As well, integral of a curve function means the area under the curve. This form of classical calculus was developed extensively over four centuries. Today, scientists are able to describe and model many physical phenomena with an ordinary differential equation [2]. In many cases, however, the classical calculus is not able to describe exactly these complex phenomena. Under a small deformation, for example, the relationship between the force and displacement in an ideal spring (small deformations in elastic materials) is linear, i.e. force is related to the zero derivative of displacement, while in an ideal damper, the force is proportional to the velocity of extension or compression. In other words, force is related to the first derivative of deformation. Which law, however, governs the materials with intermediate mechanical properties (i.e. between ideal spring and ideal damper)? With this aim, many researchers try to answer the question and to model as well as to analyse the mechanical behavior of these non-linear systems by means of fractional calculus [3].

Fractional calculus is a generalization of ordinary differentiation and integration to arbitrary non-integer order, but with this definition, many interesting questions will arise; for example, if the first derivative of a function gives you the slope of the function, what is the geometrical meaning of half derivative? In half order, which operator must be used twice to obtain the first derivative? The early history of this questions back to the birth of fractional calculus in 1695 when Gottfried Wilhelm Leibniz suggested the possibility of fractional derivatives for the first time[2] .

In this article, we aim to introduce fractional calculus as a new tool for modeling the complex systems, especially viscoelastic material. First, we briefly discuss the basic concepts of fractional calculus and explain the essential steps of the fractionalization algorithm. Next, we present an interpretation of fractional derivative and elaborate upon how fractional equations could be solved analytically. Then, we briefly look at the modeling of viscoelastic systems by the help of this approach. Ultimately, after overviewing some recent works, we present an application of the approach in modeling biomechanical properties of a cell and indicate that the proposed model predicts the cell behavior much better than the previous spring-dashpot models, as well as the model outputs are in good agreement with experimental data. To sum, we are going to give the minimum need to get reader "feet wet", so that a reader can quickly get into building a fractional calculus model for a complex system.

# Methods

# Fractional calculus

In mathematics, many complex concepts developed from simple concepts. For example, we can refer to the extension of natural number to the real one in some mathematical formulae. Let's give an example to clarify: the factorial of a non-negative integer n, denoted by n!, is the product of all positive integers less than or equal to n. On the other hand, there is a concept which named Gama function and defined as follows:

$$\Gamma(x) = \int_0^\infty t^{z-1} e^{-t} dx \qquad (1)$$

One property of the function for $n \in \mathbb{R}^+$ is

$$\Gamma(n + 1) = n\Gamma(n) \qquad (2)$$

Hence, this function is equal to factorial for the integer numbers. As a result, the gamma function could be considered as an extension of factorial function to real numbers. For instance, according to the above formalism, a factorial of $1/2$ can be obtained as follows:

$$\left(\frac{1}{2}\right)! = \Gamma\left(\frac{3}{2}\right) = \frac{3}{2}\Gamma\left(\frac{1}{2}\right) = \frac{3\sqrt{\pi}}{2} \qquad (3)$$

Indeed, according to Wikipedia, the gamma function can be seen as the solution to find a smooth curve that connects the points (x, y) given by y = (x − 1)! at the positive integer values for x. Let's use this approach to extend the concept of derivative to non-integer order; consider $n^{th}$ derivation of power function $g(x)$:

$$g(x) = x^k \ \ s.t \ \ x \geq 0 \qquad (4)$$

$$\frac{d^n}{dx^n} g(x) = \frac{k!}{(k-n)!} x^{k-n} = \frac{\Gamma(1+k)}{\Gamma(1+k-n)} x^{k-n} \qquad (5)$$

which $k$ and $n$ are real integer number respectively, and $k \geq n$. To generalize the above equation, it could be possible to extend the integer number $n$ to a real value which named $\alpha$:

$$\frac{d^\alpha}{dx^\alpha} g(x) = \frac{\Gamma(1+k)}{\Gamma(1+k-\alpha)} x^{k-\alpha} \qquad (6)$$

Then for fractional derivative of an arbitrary function, expand the function in a power series of $x$ first, and then by using $eq.\,5$, derivate the expansion. For example, for derivative $f(x) = e^{kx}$ to $\alpha$ order, we rewrite $f(x)$ function as follows:

$$f(x) = 1 + x + \frac{x^2}{2!} + \frac{x^3}{3!} + \cdots \qquad (7)$$

Hence [4]:

$$\frac{d^\alpha}{dx^\alpha} f(x) = \frac{1}{\Gamma(1-\alpha)} x^{-\alpha} + \frac{1}{\Gamma(2-\alpha)} x^{1-\alpha} + \frac{1}{\Gamma(3-\alpha)} x^{2-\alpha} + \cdots = sign(x)(sign(x)k)^\alpha e^{kx}\left(1 - \frac{\Gamma(-\alpha, kx)}{\Gamma(-\alpha)}\right) \qquad (8)$$

That $\Gamma(-\alpha, kx)$ is incomplete gamma function (discussed in Box 3).

Of course, this is an arbitrary way to define fractional derivative and not the only way; for example, it is possible to use an exponential function $f(x) = e^{kx}$ instead of a power function, i.e. we can define

$$D^\alpha f(x) = k^\alpha e^{kx} \qquad (9)$$

The fractional derivative of the exponential function obtained by Liouville in 1832, and the fractional derivative of power function got by Riemann in 1847 [4]. By comparing Eq. 8 with 9, it is noticed that when the order is an integer, the results are same, but at the non-integer order they are different! In fact, contrary to the integer-order derivative in which the definition and the output under these operators are same and unique in the fractional derivatives, under different operators, the result is not the same and not necessarily unique. In other words, there are multiple definitions for fractional derivative and all of them are mathematically correct. From the physical perspective, each definition has its own application and interpretation.

Fractional derivative has many interesting and counter-intuitive properties; for example, the Riemann derivative (Eq. 5) of a constant is not zero! that is,

$$\frac{d^\alpha}{dx^\alpha} cons = \frac{d^\alpha}{dx^\alpha} x^0 = \frac{1}{\Gamma(1-\alpha)} x^{-\alpha} \qquad (10)$$

Putting it all together, there are many ways to define a fractional derivative, provided that each definition approach to ordinary derivative in the integer order limit—this method known as the fractionalization algorithm—i.e.:

$$\lim_{\alpha \to n} \Delta^\alpha f(x) = \Delta^n f(x) \ , \ n = 0,1,....$$

Where $\Delta$ represents an arbitrary operator. As well, two more examples provided in Box 1 which have also been used in the following sections.

---

### Box 1

In this box, a basic definition of both the ordinary integral and the ordinary derivative is presented by two examples, as well as its extension to fractional operators is proposed by the help of fractionalization algorithm (ref section 2); resulting we obtain two important fractional functions.

**Example 1.** Assuming $f(x) \in [a,b]$. The first order derivative of function $f(x)$ is defined as follows:

$$f^{(1)}(t) = \frac{df}{dt} = \lim_{h \to 0} \frac{f(t) - f(t-h)}{h} \qquad (B1.1)$$

by applying derivative operator on the above equation, we get:

$$f^{(2)}(t) = \frac{d^2 f}{dt^2} = \lim_{h \to 0} \frac{f(t) - 2f(t-h) + f(t-2h)}{h^2} \qquad (B1.2)$$

by the help of induction and above procedure, we gain:

$$f^{(n)}(t) = \frac{d^n f}{dt^2} = \lim_{h \to 0} \frac{1}{h^n} \sum_{r=0}^{n} (-1)^r \binom{n}{r} f(t-rh) \qquad (B1.3)$$

Where $a \leq t \leq b$, $h = \frac{t-a}{n}$, and $\binom{n}{r} = \frac{n(n-1)(n-2)...(n-r+1)}{r!}$.

Since the equation (B1.3) hold for all $n \in N$, we expand $n$ order to $\alpha \in R$ by the fractionalization algorithm; thereby achieving a really important equation known as fractional Grunwald-Letnikov differintegral mentioned in section 2.

$$f_n^{(\alpha)}(t) = \lim_{h \to 0} \frac{1}{h^\alpha} \sum_{r=0}^{\infty} (-1)^r \binom{\alpha}{r} f(t-rh) \qquad (B1.4)$$

**Example 2.** Consider the following equation which is known as the integral operator of order n:
We can rewrite it as follow according to the Cauchy equation [6]:

$$_\alpha D^{-n} f = \int_\alpha^{x_n} \int_\alpha^{x_{n-1}} \cdots \int_\alpha^{x_1} f(x_0) dx_0 \cdots dx_{n-1} \qquad (B1.5)$$

Since the above equation holds for all $n \in N$, we apply the fractionalization algorithm to expand the obtained equation to fractional equation. In other words, if n → α, we will able to gain an integral known as Riemann-Liouville. Hence, it seems that fractional calculus is an expansion of integer calculations.

$$_\alpha D^{-n} f = \frac{1}{(n-1)!} \int_\alpha^x (x-\tau)^{n-1} f(\tau) d\tau \qquad (B1.6)$$

# Definition of fractional calculus

As mentioned in the previous section, the fractional order derivative is not necessarily unique; in this regard, there are some accepted and common definitions in the literature. In the following, we mentioned a number of important ones.

**Grünwald–Letnikov derivative**

Grünwald–Letnikov derivative is a basic extension of the natural derivative to fractional one, which derived in Box 1 (Eq. B1.4). It was introduced by Anton Karl Grünwald in 1867, and then by Aleksey Vasilievich Letnikov in 1868. Hence, it is written as [5]:

$$D^\alpha f(t) = \lim_{h \to 0} \frac{1}{h^\alpha} \sum_{m=0}^{\infty} (-1)^m \binom{\alpha}{m} f(t - mh) \tag{11}$$

where $n \in N$, and the binomial coefficient is calculated by the help of the Gamma function.

$$\binom{\alpha}{m} = \frac{\alpha(\alpha-1)(\alpha-2)...(\alpha-m+1)}{m!} \tag{12}$$

**Riemann-Liouville fractional derivative**

Riemann-Liouville fractional derivative acquiring by Riemann in 1847 is defined as follows.

$$\begin{aligned}{}_a^{RL}D_t^\alpha f(t) &= (\frac{d}{dt})^n ({}_aD_t^{-(n-\alpha)}) f(t) \\ &= \frac{1}{\Gamma(n-\alpha)} (\frac{d}{dt})^n \int_a^t \frac{f(x)}{(t-x)^{\alpha-n+1}} dx, (n = [\alpha]+1, t > a)\end{aligned} \tag{13}$$

where $\alpha > 0$; this operator is an extension of Cauchy's integral (Box 1) from the natural number to real one. Based on the fractionalization algorithm, it seems logical to reach relation (13) by n order derivative of the Eq. B1.6

In addition, according to the above relation, if $0 < \alpha < 1$ then the Riemann-Liouville operator reduced to

$${}_a^{RL}D_t^\alpha f(t) = \frac{1}{\Gamma(1-\alpha)} \frac{d}{dt} \int_a^t \frac{f(x)}{(t-x)^\alpha} dx \tag{14}$$

It is worth noting that this relation is the same of Eq. 5. [5], and also RL, $\alpha$, $a$, and $t$ are the abbreviation of Riemann-Liouville, fractional order, the lower and upper bound of the above integral respectively.

**Caputo derivative**

Since Riemann-Liouville fractional derivatives failed in the description and modeling of some complex phenomena, Caputo derivative was introduced in 1967 [6]. The Caputo derivative of fractional order α ($n - 1 \leq \alpha < n$) of function $f(t)$ defined as

$${}_a^CD_t^\alpha f(t) = \frac{1}{\Gamma(n-\alpha)} \int_a^t \frac{D^n f(\tau)}{(t-\tau)^{\alpha-n+1}} d\tau \tag{15}$$

where $D^n$ is $n^{th}$ derivative operation, and C represents the Caputo word. In other words, based on box 2, it can be found that the Caputo derivative is equal to the Riemann–Liouville integral of a $n^{th}$ derivative of a function.

It is worth noting again that "the behavior of all of the fractional derivatives, when the order is integer are same; it, however, could be different in the non-integer order; for example, in the non-integer order, the Caputo derivative of a constant, unlike the Riemann–Liouville derivative is zero.

There are more fractional derivatives for more detailed information, the interested reader is referred to [4, 5, 7]. Since we try to analyse and to investigate viscoelastic systems via the Riemann-Liouville and Caputo fractional derivative, the main focus of the article is on these two types of derivative.

As previously mentioned, different definitions for fractional derivative with the different properties can be proposed, which all of them are valid and mathematically acceptable. However, the main question is "which relation should be applied in modeling of a specific phenomenon? In other words, which definition would be more appropriate for a specific problem?" As a rule of thumb, since they tend to interpret natural phenomena, the definition which is more consistent with the experimental results have more privilege than the other fractional definitions.

## BOX 2

In this box, some properties of fractional Riemann-Liouville differintegral, such as commutative property, distributive laws, and so forth are investigated[2].

**Riemann–Liouville integral and derivative**

To find a profound understanding of Riemann–Liouville integral and derivative, some of the most crucial properties of this operator are mentioned in the following.

**Lemma 1.** Assuming arbitrary function $f(x)$ and $m, n \geq 0$ the following equations hold[3].

1. Semi-group property:
$$I_a^m I_a^n f = I_a^{m+n} f$$

2. commutative property:
$$I_a^m I_a^n f(x) = I_a^n I_a^m f(x)$$

**Lemma 2.** Let $f_1$ and $f_2$ are two functions on [a,b] as well as $c_1, c_2 \in \mathbb{R}$, $n > 0$, and $m > n$. Regarding these, the following equations hold[4]:

1. Linearity rules:
$$^{RL}D_a^n(f_1 + f_2) = {^{RL}D_a^n f_1} + {^{RL}D_a^n f_2} \quad , \quad {^{RL}D_a^n(c_1 f_1)} = c_1 {^{RL}D_a^n(f_1)}$$

2. Zero rule:
$$D^0 f = f$$

3. Product rule:
$$^{RL}D_t^q(fg) = \sum_{j=0}^{\infty} \binom{q}{j} {^{RL}D_t^j(f)} {^{RL}D_t^{q-j}(g)}$$

4. In the general, semi-group property does not hold for Riemann-Liouville fractional derivative. Indeed, the following equation is not always true.
$$^{RL}D^a {^{RL}D^b} f = {^{RL}D^{a+b}} f$$

---

[2] All the mentioned properties in this box hold almost everywhere (a.e) on [a, b] and also, if (X, Σ, μ) is a measure space, a quality P is said to hold almost everywhere in X if $\mu(\{x \in X: P(x)\}) = 0$. Also, integral operator on interval $[a, b]$ is defined as $I_a^b$.

[3] $f \in L_1[a, b]$ where $L_1 \coloneqq \{f: [a, b] \to \mathbb{R}; f \text{ is measurable on } [a, b] \text{ and } \int_b^a |f(x)|^1 dx < \infty \}$.

[4] Provided that $D_a^n f_1$ and $D_a^n f_2$ almost everywhere define.

**NB:** To prove the above lemmas see [8].

**Caputo derivative**

In this part, some fundamental properties related to Caputo operator are represented based on ref [5, 8].

Let $f$ is a enough differentiable function, $c_1, c_2 \in \mathbb{R}$, and $m > n \geq 0$. Considering these,
1. Caputo derivative is the left inverse of Riemann-Liouville integral.
$$^{C}D_a^n I_a^n f = f$$
2.
$$I_a^{n\,C}D_a^n f(x) = f(x) - \sum_{k=0}^{m-1} \frac{D^k f(a)}{k!}(x-a)^k$$
3. Distributive law in Caputo derivative.
$$^{C}D_a^n(c_1 f_1 + c_2 f_2) = c_1\left(^{C}D_a^n f_1\right) + c_2\left(^{C}D_a^n f_2\right)$$
4. Leibniz equation[5].
$$^{C}D_a^n[fg](x) = \frac{(x-a)^{-n}}{\Gamma(1-n)} g(a)(f(x)-f(a)) + \left(^{C}D_a^n g(x)\right)f(x) + \sum_{k=1}^{\infty} \binom{n}{k}\left(I_a^{k-n}g(x)\right){}^{C}D_a^k f(x)$$
5. The semi-group property, the following equation, holds under some special condition[6].
$$^{C}D_a^\alpha\,^{C}D_a^\beta f = {}^{C}D_a^{\alpha+\beta}f$$

**NB**: The part 5 of the above Lemma) does not hold for Riemann–Liouville derivative generally. To prove this claim, assuming $f$ function with identity function $f(x) = 1$, as well as $a = 0, n = 1$, and $\varepsilon = 1/2$. Thus, if Riemann–Liouville operator is applied to the left side, we will obtain $(D_0^{\frac{1}{2}}f')(x) = D_0^{\frac{1}{2}}0 = 0$. On the other hand, by applying this operator to the right side, we will gain

$$D_0^{3/2}f(x) = D^2 I_0^{1/2}f(x) = \frac{1}{\Gamma(-1/2)}x^{-3/2}$$

**The relationship between fractional integral and fractional derivative**

In this part, the relationship between fractional integral and fractional derivative is investigated. There are always several properties in the classical integral and derivative area accepted as constant roles; these properties, however, might not hold always in the fractional sense. The following equations, for instance, are almost same in the classical calculation, but it has a major difference in the fractional calculation $(m > \alpha)$.

$$D^\alpha = D^{\alpha-m}D^m = {}_aI^{m-\alpha}\frac{d^m}{dx^m} \quad \text{and} \quad D^\alpha = D^m D^{\alpha-m} = \frac{d^m}{dx^m}{}_aI^{m-\alpha}$$

Indeed, a constant and a variable term in calculation represent this difference, which the constant and variable term are related to non-fractional and fractional sense respectively. In other words, assuming function f: [a, b] → ℝ where $-\infty < a < b < \infty$; if the function $f(x)$ on closed interval $[a, b]$ has been $n$ order differentiable, then for all $x \in (a, b)$, we have:

---

[5] Assuming $0 < n < 1, g$ and $f$ are analytical on $(a-h, a+h)$ are essential to hold the equation.

[6] Special condition for existing this property is $f$ function must belong to $C^k[a,b]$ where $C^k[a,b] := \{f:[a,b] \to \mathbb{R}; f \text{ has a continuous } Kth \text{ derivative}\}$ for some $a < b$ and some $k \in \mathbb{N}$. Existing $\ell \in \mathbb{N}$ with $\ell \leq k$ and $n, n + \varepsilon \in [\ell-1, \ell]$ is essential as well $(n, \varepsilon > 0)$.

**NB:** The existence of $\ell$ with the mentioned properties is essential. Otherwise, it is enough to assume that $n = \varepsilon = 7/10 (i.e\ 7/10 = n < 1 < n + \varepsilon = 7/5)$, $a = 0$, and $f(x) = x$, causing the right side of lemma become zero owing to $^{C}D_0^{7/5}f(x) = (I_0^{3/5}f'')(x) = I_0^{3/5}0 = 0$; however, $^{C}D_0^{7/10}f(x) = \frac{1}{\Gamma(13/10)}x^{13/10}$. As a result, the left side of lemma will obtain as follow:

$$^{C}D_0^{7/10}(^{C}D_0^{7/10}f)(x) = \frac{1}{\Gamma(3/5)}x^{-2/5}$$

$$(^C D_a^\alpha I_a^\alpha f)(x) = f(x)$$

In a simple word, if the fractional derivative with order $\alpha \in \mathbb{R}$ is applied on the fractional integral with the same order, the output of that will be $f(x)$. Thus, we can say, "Caputo derivative is the left inverse of Riemann-Liouville integral." Also

$$(I_a^\alpha \, {}^C D_a^\alpha f)(x) = f(x) - \sum_{j=0}^{n-1} \frac{|x-a|^j}{j!} \left(f^{(j)}(x)\Big|_{x=a}\right)$$

**NB**: To prove the mentioned theorems see [9].

With regarding the Riemann-Liouville as a derivative operator, the following equation hold ($n > 0$, $m > n$):

$$I_a^n \, {}^{RL}D_a^n f(x) = f(x) - \sum_{k=0}^{m-1} \frac{(x-a)^{n-k-1}}{\Gamma(n-k)} \lim_{z \to a+} D^{m-k-1} I_a^{m-n} f(z)$$

and particularly, for $0 < n < 1$, we have:

$$I_a^n \, {}^{RL}D_a^n f(x) = f(x) - \frac{(x-a)^{n-1}}{\Gamma(n)} \lim_{z \to a+} I_a^{1-n} f(z)$$

# The interpretation of fractional calculus

There is a dozen of suggestions for interpretation of fractional calculus [4, 7]. However, most of them are a little bit abstract and give no physical intuition. Hence, we just present a most useful interpretation of fractional calculus. The core of these interpretations is memory concept. In general, when the output of a system at each time $t$ depends only on the input at time $t$, such systems are said to be memoryless systems. On the other side, when the system has to remember previous values of the input in order to determine the current value of the output, such systems are said non-memoryless systems, or memory systems. For example, all of the Markov chain phenomena are memoryless [3, 4], and human decision making or shape memory alloy are non-memoryless.

## Laplacean interpretation

Suppose that $Y(t)$ is a quantity whose value in terms of $f(t)$ can be achieved as follows:

$$Y(t) = \int_0^t \frac{(t-\tau)^{(\alpha-1)}}{\Gamma(\alpha)} f(\tau) d\tau \tag{16}$$

i.e. the output $Y(t)$ can be viewed as a power-weighted sum which stores the previous input of function $f(t)$. Based on the above definition, such system is a non-memoryless system and in such systems, memory decays at the rate of $w(t) = t^{\alpha-1}/\Gamma(\alpha)$.

Applying the Caputo derivative of order $\alpha$ to the both sides of the last relation leads to

$$^C_0 D_t^\alpha Y(t) = f(t) \tag{17}$$

As a result, the differential equation governing the system memory $Y(t)$ is described by a fractional derivative. Therefore, the fractional derivative is a good candidate to explain the system with memory.

The nature of weighted function determines the type of fractional derivative which describes a system memory. For example, If the weight function of a system is defined by $t^{1-\alpha}/\Gamma(\alpha)$, the Riemann-Liouville elements, and by $t^{1-\alpha}\theta(x-t)/\Gamma(\alpha)$ the Caputo elements are used--which θ is the Heaviside function [4].

Since the convolution operation representing the integral of the product of the two functions after one is reversed and shifted, so if one of the functions considered as a weighted function, it may be claimed that the task of this function is collecting the system memory over time. For example, in the Riemann-Liouville, "The relationship indicates that the information function $F(t)$ is memorized (storage) with a power low rate function". In Ref [4] some other weight function is reviewed.

From the physical point of view, what the memory is and how it is defined in a system depends on a deep understanding of the phenomena. Also, there is no certain rules and methods to select the fractional type in a modeling. In other words, any definition that its result is more consistent with the experimental data is the suitable definition to us.

# Analytical and numerical methods

## Analytical methods

There are different methods to solve fractional differential equations analytically. One of the most common and widely used methods is the Laplace transform. In the following, by an example, this method is illustrated (for other methods see reference [7]).

Before proceeding, it is worth noting that in general, the number of initial conditions that are required for a given ordinary differential equation will depend upon the order of the differential equation. However, in the fractional differential equation, number of initial condition equals to the integer lower bound of order value ($\alpha$) [10, 11].

Consider the following differential equation

$$\xi D^{\alpha} x(t) + k x(t) = f \qquad (18)$$

Which $x(t)$ is displacement, $k$, and $\xi$ are constants, as well as the fractional derivative is also Caputo and $0 < \alpha < 1$--In what follows, it has been shown that this differential equation model is the dynamics of a purely elastic spring and a viscoelastic element connecting in parallel with a body of mass m, which a force f is applied on a body. In order to solve, the first step is to take the Laplace transform of both sides of the original differential equation (the Laplace transformation is concisely explained in Box 3). Based on the equation B3.2 in Box 3, we have

$$X(s) = \frac{f}{\xi(s^{\alpha} + k/\xi)} \qquad (19)$$

where $\alpha$ and $s$ are fractional order and Laplace domain variable respectively. Also, it is supposed that $x(0) = 0$. To find the solution, all we need to do is to take the inverse transform (Box 3.1):

$$x(t) = \frac{f}{\xi} t^{\alpha-1} E_{\alpha,\alpha}(-\frac{k}{\xi} t^{\alpha}) \qquad (20)$$

Which $E_{\alpha,\alpha}(t)$ is Mittag-Leffler function [for more information see the Box 3 footnote]

In the last example, if the spring is ignored, the Eq. 18 will be reduced to
$$f = \xi D^\alpha x(t) \quad (21)$$

Then as above, by a) taking the Laplace transforms of both sides of the equation, b) simplifying algebraically the result to solve the obtained equation in terms of $s$, and c) finally finding the inverse transform, we have:
$$x(t) = Kt^\alpha \quad (22)$$

Which $K = f/\xi\Gamma(1 + \alpha)$. Although the Laplace transformation method is one of the simple and practical methods for solving the fractional equations--same as the ordinary differential equations, most of the fractional equation could not be solved analytically. In what follows, we present a numerical technique to solve Caputo fractional differential equation.

## BOX 3

In this box, the concept of Laplace Transform is presented for several fractional definitions. To begin with, let us bring up some basic facts in this regard [7].

The function $F(s)$ of the complex variable $s$ is defined as the following equation—which is known as the Laplace transform of the function $f(t)$:
$$F(s) = L\{f(t)\} = \int_0^\infty e^{-st} f(t)\,dt \quad (B3.1)$$

To exist integral (B3.1), the function $f(t)$ must have been an exponential order $\alpha$. In other words, the existence of two positive constants, such as $M$ and $T$ which satisfy the following condition is essential.
$$e^{-\alpha t}|f(t)| \leq M \quad \text{for all } t > T.$$
Indeed, when $t \to \infty$, the function $f(t)$ cannot grow faster than a certain exponential function.

With the help of the inverse Laplace transform, the original $f(t)$ can be gained from the Laplace transform $F(s)$
$$f(t) = L^{-1}\{F(s)\} = \frac{1}{2\pi i} \lim_{b \to \infty} \int_{a-ib}^{a+ib} e^{st} F(s)\,ds \quad (B3.2)$$

where the integration is done along the vertical line $Re(s) = a$ in the complex plane such that $a$ is greater than the real part of all singularities of $F(s)$, which guarantees that the contour path is in the convergent region. If either all singularities are in the left half-plane or $F(s)$ is a smooth function on $-\infty < Re(s) < \infty$ (i.e., no singularities), then $a$ can be set to zero and the above inverse integral formula becomes identical to the inverse Fourier transform.

The direct calculation of the inverse Laplace transform (B3.2) is often sophisticated. Sometimes, however, it gives useful information on the unknown original $f(t)$ behavior which we look for.

The following formula seems to be another useful property for the Laplace transform of the derivative of an integer order $n$ of the function $f(t)$:
$$L\{f^{(n)}(t)\} = s^n F(s) - s^{n-1} f(0) - s^{n-2} f'(0) - \cdots - f^{(n-1)}(0) = s^n F(s) - \sum_{k=0}^{n-1} s^{n-k-1} f^{(k)}(0) \quad (B3.3)$$

which can be obtained from the definition (B3.1). To facilitate applying the Laplace transform, we gather some formulae of vital functions in the table (B3.1) for both the integer mode and the non-integer mode.

**Table B3.1.** Laplace transform table of some basic fractional calculus.

| Fractional order | | Integer order | |
|---|---|---|---|
| $f(t) = L^{-1}\{F(s)\}$ | $F(s) = L\{f(t)\}$ | $f(t) = L^{-1}\{F(s)\}$ | $F(s) = L\{f(t)\}$ |
| $\dfrac{1}{s^{\alpha}}$ | $\dfrac{t^{\alpha-1}}{\Gamma(\alpha)}$ | $1$ | $\delta(t)$ |
| $\dfrac{1}{(s+a)^{\alpha}}$ | $\dfrac{t^{\alpha-1}}{\Gamma(\alpha)}e^{-at}$ | $\dfrac{1}{s^{k}}$ | $\dfrac{t^{k-1}}{\Gamma(k)},\ k=1,2,3,\ldots$ |
| $\dfrac{a^{\alpha}}{s(s+a)^{\alpha}}$ | $\dfrac{1}{\Gamma(\alpha)}\gamma(\alpha,at)$ [7] | $aF(s)+bG(s)$ | $af(t)+bg(t);\ a,b\in\mathbb{R}$ |
| $\dfrac{s^{\alpha}}{s(s^{\alpha}+a)}$ | $E_{\alpha}(-at^{\alpha})$ [8] | $s^{n}L\{f(t)\}-s^{n-1}f(0)$ $-s^{n-2}f'(0)-\cdots-f^{(n-1)}(0)$ | $f^{(n)}(t),\ n>0$ |
| $\dfrac{1}{s^{\alpha}+a}$ | $t^{\alpha-1}E_{\alpha,\alpha}(-at^{\alpha})$ | $\dfrac{a}{s^{2}+a^{2}}$ | $\sin(at)$ |
| $\dfrac{a}{s(s^{\alpha}+a)}$ | $1-E_{\alpha}(-at^{\alpha})$ | $\dfrac{s}{s^{2}+a^{2}}$ | $\cos(at)$ |
| $\dfrac{s^{\alpha}}{s(s-a)^{\alpha}}$ | ${}_1F_1(\alpha;1;at)$ [9] | $F(s-c)$ | $e^{ct}f(t)$ |
| $\dfrac{1}{s^{\alpha}(s-a)}$ | $t^{\alpha}E_{1,1+\alpha}(at)$ | $(-1)^{n}F^{(n)}(s)$ | $t^{n}f(t),\ n=1,2,3,\ldots$ |
| $\dfrac{s^{\alpha}}{(s-a)}$ | $t^{-\alpha}E_{1,1-\alpha}(at),\ 0<\alpha<1$ | $F(s)G(s)$ | $\int_{0}^{t}f(t-\tau)g(\tau)d\tau$ |
| $\dfrac{s^{\alpha-\beta}}{s^{\alpha}-a}$ | $t^{\beta-1}E_{\alpha,\beta}(at^{\alpha})$ | $\int_{s}^{\infty}F(u)du$ | $\dfrac{1}{t}f(t)$ |
| $\dfrac{s^{\alpha-\beta}}{(s-a)^{\alpha}}$ | $\dfrac{t^{\beta-1}}{\Gamma(\beta)}{}_1F_1(\alpha;\beta;at)$ | $\dfrac{F(s)}{s}$ | $\int_{0}^{t}f(v)dv$ |

## The Laplace transform of fractional operators

In this part, the Laplace transform of fractional operators is represented, and some formulae of which is summarized in table (B3.2).

We introduced Riemann-Liouville integral in Box 1 as follow:

$$_0I_t^{\alpha}f = \frac{1}{\Gamma(\alpha)}\int_0^t (t-\tau)^{\alpha-1}f(\tau)d\tau = \frac{1}{\Gamma(\alpha)}[t^{\alpha-1}*f(t)] \quad (B3.4)$$

With applying the Laplace transform on the Eq. B3.4, we obtain (based on Table B3.1):

$$L\{_0I^{\alpha}f\} = \frac{1}{\Gamma(\alpha)}L\{t^{\alpha-1}\}L\{f(t)\} = s^{-\alpha}F(s)$$

---

[7] incomplete gamma function

$$\Gamma(a,z) = \int_z^{\infty} t^{a-1}e^{-t}dt$$

[8] $E_{\alpha,\beta}$ is denoted to Mittage-Leffler function such that

$$E_{\alpha,\beta}(x) = \sum_{k=0}^{\infty}\frac{x^k}{\Gamma(k\alpha+\beta)},\quad \alpha,\beta>0$$

Also, if $\alpha = \beta = 1$, then this function can be equaled with exponential function.

[9] Hypergeometric functions

$$_pF_q(\{a_i\};\{b_j\};z) = \frac{\prod_{j=1}^{q}\Gamma(b_j)}{\prod_{i=1}^{p}\Gamma(a_i)}\sum_{n=0}^{\infty}\frac{\prod_{i=1}^{p}\Gamma(a_i+n)}{\prod_{j=1}^{q}\Gamma(b_j+n)}\frac{z^n}{n!}$$

Similarly, other Laplace transform of fractional operators will be gained by utilizing the proposed formulae in Table B3.1 [7].

**Table B3.2.** The Laplace transform of some fractional operators with order $\alpha$.

| Operator | Fractional formula | Laplace Transform of the fractional order operators |
|---|---|---|
| Riemann-Liouville integral | $_0I_t^\alpha f(t) = \frac{1}{\Gamma(\alpha)}\int_0^t (t-\tau)^{\alpha-1} f(\tau)d\tau$ | $s^{-\alpha}F(s)$ |
| Riemann-Liouville derivative | $^{RL}_0D_t^\alpha f(t) = \frac{1}{\Gamma(n-\alpha)}(\frac{d}{dt})^n \int_a^t \frac{f(x)}{(t-x)^{\alpha-n+1}}dx$ | $s^\alpha F(s) - \sum_{k=0}^{n-1} s^k \left[^{RL}_0D_t^{\alpha-k-1}f(t)\right]_{t=0}$, $n-1 \leq \alpha < n$ |
| Caputo derivative | $^C_0D_t^\alpha f(t) = \frac{1}{\Gamma(n-\alpha)}\int_a^t \frac{D^n f(\tau)}{(t-\tau)^{\alpha-n+1}}d\tau$ | $s^\alpha F(s) - \sum_{k=0}^{n-1} s^{\alpha-k-1}f^{(k)}(0)$, $n-1 < \alpha \leq n$ |
| Grünwald-Leitnikov fractional derivative[10] | $_0D_t^\alpha f(t) = \lim_{\substack{h\to 0 \\ nh=t-a}} h^{-\alpha} \sum_{m=0}^n (-1)^m \binom{\alpha}{m} f(t-mh)$ | $s^\alpha F(s)$, $0 < \alpha < 1$ |

## Numerical solution

Until now, various numerical methods have been proposed for solving fractional differential equations which are based on discretizing the relationships [12-17]. In this section, a new algorithm which was recently published in [18] is proposed discretizing the Caputo derivative.

Considering the Caputo derivative of function $u(t)$ with order $0 < \alpha < 1$, the following equations hold for discretizing $u(t)$ on [0,T] in a uniform way.

$$\partial_t^\alpha u(t_{n+1}) = \frac{1}{\Gamma(1-\alpha)} \sum_{j=0}^n \int_{t_j}^{t_{j+1}} (t_{n+1}-s)^{-\alpha} \frac{\partial u(s)}{\partial s} ds$$
$$= \frac{1}{\Gamma(1-\alpha)} \sum_{j=0}^n \frac{u(t_{j+1})-u(t_j)}{\tau} \int_{t_j}^{t_{j+1}} (t_{n+1}-s)^{-\alpha} ds + r_{\alpha,\tau}^{n+1} \quad (23)$$
$$= \frac{1}{\Gamma(2-\alpha)} \sum_{j=0}^n d_{\alpha,j} \frac{u(t_{n+1-j})-u(t_{n-j})}{\tau^\alpha} + r_{\alpha,\tau}^{n+1}$$

Where $d_{\alpha,j} = (j+1)^{1-\alpha} - j^{1-\alpha}$ and $j = 0,1,2,...,n$, as well as $r_{\alpha,\tau}^{n+1}$ is local truncation error.

To utilize the above algorithm, equation (21), which used in the last section simulation, is considered. The following equation will be obtained by applying the equation (23) on equation (21).

$$x(t_{n+1}) = \tau^\alpha \Gamma(2-\alpha)(f/\xi) + x(t_n) - \sum_{j=1}^n d_{\alpha,j}(x(t_{n+1-j}) - x(t_{n-j})) \quad (24)$$

Comparison between the numerical method and the analytical answer is shown in both the figure (1) and the table (1)

---

[10] The Laplace transform of the Grünwald-Letnikov fractional derivative of order $\alpha > 1$ does not exist in the classical sense.

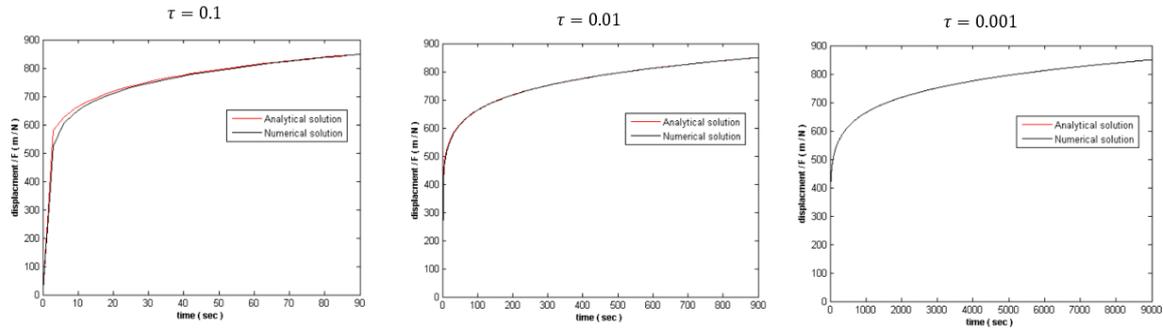

Fig 1. A schematic comparison of analytical and numerical answer with the represented mesh in Table 1.

**Table 1.** Comparison between the numerical results, equation 24, and the analytical answers of equation (25) with $f = 1100 pN$ stress ($\xi$ and $\alpha$ are chosen from table 4).

| Mesh point ($\tau$) | Numerical result in $t = 3$ | Analytical result in $t = 3$ | Difference | Percentage error |
|---|---|---|---|---|
| 0.1 | 848.3334 | 849.9605 | 1.6271 | 0.1914 |
| 0.01 | 849.8002 | 849.9605 | 0.1603 | 0.0188 |
| 0.001 | 849.9445 | 849.9605 | 0.0160 | 0.0018 |

# Viscoelastic systems

Elasticity is the ability of a material to resist a distorting or deforming force and return to its original form when the force is removed. According to the classical theory in the infinitesimal deformation, most elastic materials, based on Hooke's Law (Eq. 25), can be described by a linear relation between the stress $\sigma$ and strain $\varepsilon$ [19].

$$\sigma(t) = E\varepsilon(t) \qquad (25)$$

where $E$ is a constant known as the elastic or Young's modulus.

The viscosity of a fluid is a quantity which describes its resistance to deformation under shear stress or tensile stress. In the ideal viscose fluid, according to the Newtonian fluid (Eq. 26) the stress is proportional to the local shear velocity (i.e. the rate of deformation over time):

$$\sigma(t) = \eta \frac{d\varepsilon(t)}{dt} \qquad (26)$$

where $\eta$ is the shear modulus.

However, many liquids briefly respond like elastic solids when subjected to abrupt stress. On the contrary, many solids will flow like liquids, albeit very slowly, under small stress. Such materials possessing both elasticity (reaction to deformation) and viscosity (reaction to the rate of deformation) are known as viscoelastic materials. When these materials subjected to sinusoidal stress, the corresponding strain is neither in the same phase as the applied stress (like an ideal elastic material) nor in the $\pi/2$ out of phase (like an ideal viscose material). In these materials, some part of input energy is stored and recover in each cycle and the other part of the energy is dissipated as heat. Materials whose behavior exhibits these characterized are called viscoelastic. If both strain and stress be not only infinitesimal but dependent upon the time as well, strain-stress relation (constitutive equation) can be described by a linear differential equation with constant coefficient i.e. the material which exhibits linear viscoelastic behavior. One of the main consequent of this assumption is the stress (strain) responses to successive strain (stress) stimuli are additive. In other words, in the creep experiment consist of applying a step stress $\sigma_0$, (a stress increment at time t = 0, which is kept constant for t>0), and measuring the corresponding stain respond $\varepsilon(t)$, the constitutive equation is $\varepsilon(t) = \sigma_0 J(t)$, where $J(t)$—which termed creep compliance—is the strain at time t owing to a unit stress increment at time

0. Suppose that various stress increments $\Delta \sigma$ occur at the successive time interval $\Delta t$ based on the superposition of effects, we get:

$$\varepsilon(t) = \sigma_0 J(t) + \Delta \sigma' J(t-t') \qquad (27)$$

After N steps, we have

$$\varepsilon(t) = \sigma_0 J(t) + \sum_{n=1}^{N} J(t - n\Delta t') \left(\frac{\Delta \sigma}{\Delta t}\right)_{t=n\Delta t'} \Delta t \qquad (28)$$

This relation recognized as Boltzmann superposition principle which states that the response of a material to a given load is independent of the response of the material to any load, which is already on the material [19].

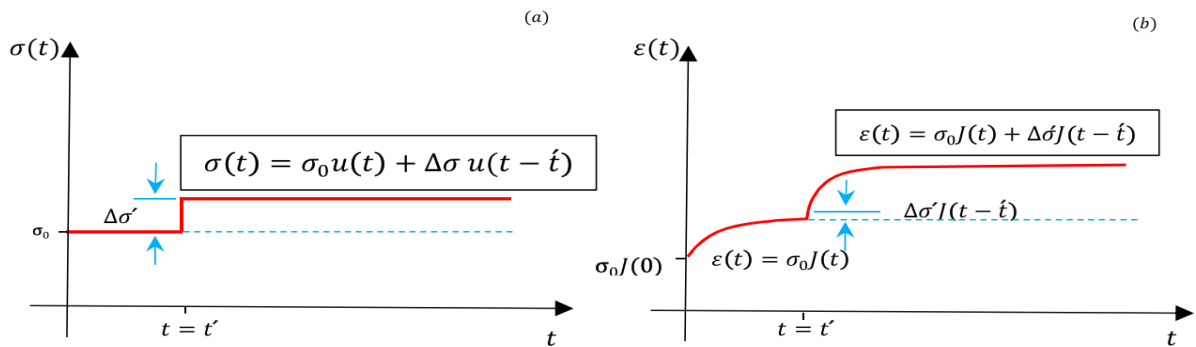

**Fig 2.** A graph versus time of (a) a step-plus increment stress applied to linear viscoelastic material and (b) the resulting strain.

in the limit $\Delta t \to 0$

$$\varepsilon(t) = \sigma_0 J(t) + \int_0^t J(t-t') \frac{d\sigma(t')}{dt'} dt' \qquad (29)$$

This relation is termed hereditary or superposition integral. Similarly, with applying a step strain, we get

$$\sigma(t) = \varepsilon_0 G(t) + \int_0^t G(t-t') \frac{d\varepsilon(t')}{dt'} dt' \qquad (30)$$

Where G(t) is relaxation modulus.

In the mechanical science, to evaluate mechanical properties of materials, after sudden stress and dynamic (periodic) experiments, researchers implement certain tests, such as creep tests. In the following, we will briefly discuss these two tests.

## Creep test

One of the most common tests on the viscoelastic system is creep test. In this test, a system is subjected to constant tension and the amount of deformation is recorded over time. For viscoelastic materials creep vs. time diagram shows power-law behavior (figure 3).

$$\varepsilon(t) = \sigma J(t) \qquad (31)$$

Where $J(t) = 1/G(t)$ is termed creep compliance.

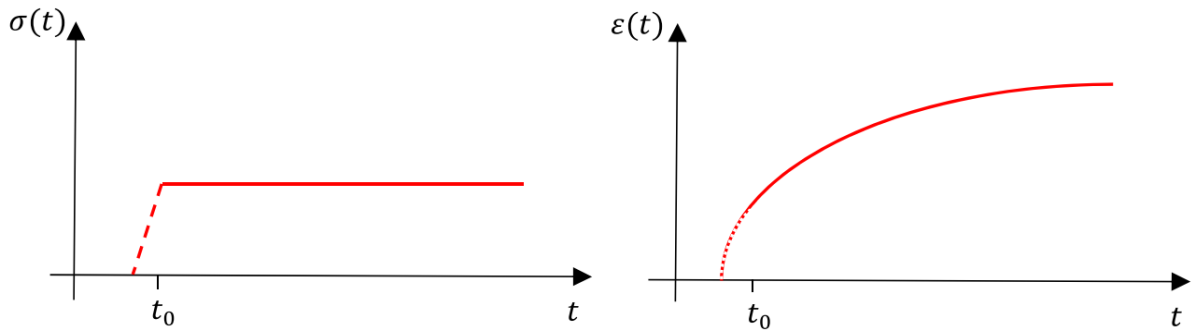

**Fig 3.** The profile of a shear creep experimental.

## Periodic test

In this common experiment, a system subjected to a cyclic applied stress. Therefore, as mentioned above, for a viscoelastic system, strain in response to an imposed periodic stress of angular frequency $\omega$ is also periodic with the same frequency but in the different phase (figure 4). Based on relation (30), if the periodic stress $\varepsilon = \varepsilon^0 \sin(\omega t)$ applied, we expect that the stress follows a periodic response with the same frequency and different phase:

$$\sigma = \sigma^0 \cos(\omega t + \delta) = \sigma^0 \cos\delta \sin\omega t + \sigma^0 \sin\delta \cos\omega t \qquad (32)$$

And by extension of relaxation modulus, we define two shear storage modulus G', and shear loss modulus G" as follows

$$\begin{aligned} G' &= (\sigma^0/\varepsilon^0)\cos\delta \\ G'' &= (\sigma^0/\varepsilon^0)\sin\delta \\ G''/G' &= \tan\delta \end{aligned} \qquad (33)$$

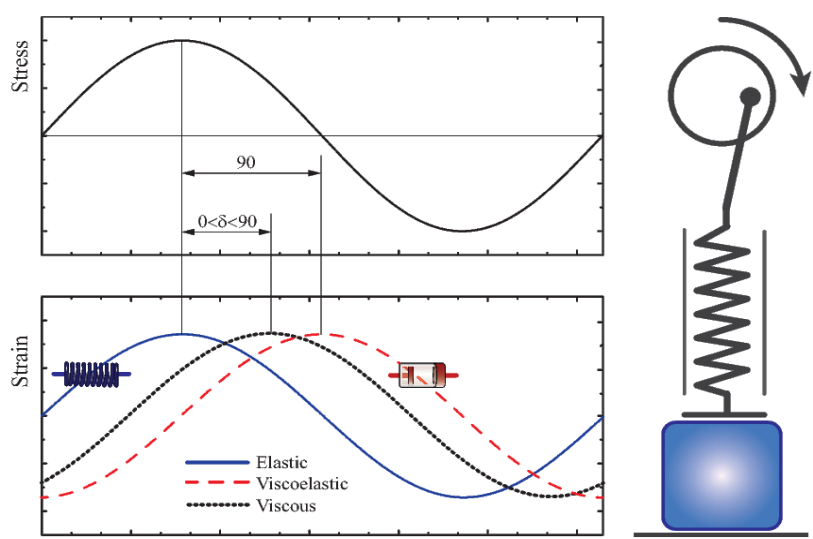

**Fig 4.** Geometry and time profile of a simple shear experiment with sinusoidally varying shear [19].

# Fractional viscoelastic model

There are different models to evaluate and to predict the constitutive equation in the viscoelastic system, these models commonly, composing of a different combination of springs and dampers elements [20, 21]. For example, two well-known models to name are Maxwell and Kelvin models (Table 2).

In this section, a model which is known as fractional order element (FOE) with equation (39) is presented in order to investigate the mechanical response of viscoelastic systems from different points of view. It is indicated that this fractional model can be a good substitute for modeling the viscoelastic materials than SLE models. With this aim, let us take a closer look at Maxwell model; based on this model, the constitutive equation will be as follows:

$$\frac{1}{E}\frac{d\sigma}{dt} + \frac{\sigma}{\eta} = \frac{d\varepsilon}{dt} \tag{34}$$

where $E(0^+) = \eta(0^+)$. Also, $\sigma, \varepsilon, \eta$, and $E$ are stress, strain, shear modulus, and Young modulus respectively. The following equation will be obtained by taking Laplace transformation and considering a step strain function.

$$G(t) = Ee^{-\left(\frac{t}{\tau}\right)} \tag{35}$$

Where $t > 0$ and $G(t)$ is relaxation modulus.

In nature, the $G(t)$ value in viscoelastic materials does not usually follow a simple exponential behavior, and a more realistic expression for its behavior is a power-low response [3]:

$$G(t) = \frac{E}{\Gamma(1-\alpha)}\left(\frac{t}{\tau}\right)^{-\alpha} \tag{36}$$

**Table 2.** Viscoelastic models, which $\sigma, \varepsilon\ and\ \tau$ are stress, strain and $\eta/E$ respectively; also, $X_\alpha = E^{1-\alpha}\eta^\alpha$ and $0 < \alpha < 1$.

| Name | Model | Constitutive equation | Relaxation modulus ($G(t)$) |
|---|---|---|---|
| **Maxwell fluid** | 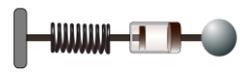 | $\sigma(t) + \tau D\sigma(t) = \eta ED\varepsilon(t)$ | $Ee^{-t/\tau}$ |
| **Voigt fluid** | 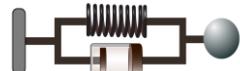 | $\sigma(t) = E\varepsilon(t) + \eta D\varepsilon(t)$ | $E + \eta\delta(t)$ |
| **FOE** | 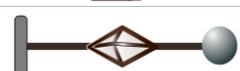 | $\sigma(t) = X_\alpha D^\alpha \varepsilon(t)$ | $\frac{E}{\Gamma(1-\alpha)}\left(\frac{t}{\tau}\right)^{-\alpha}$ |

As it mentioned in the previous section, the value of stress in a pure elastic system, an ideal spring, is proportional to the derivative zero displacements. On the other hand, it is proportional to the first-order derivative of displacement in the pure viscous system. It is expected that the viscoelastic systems have a behavior between these two systems, an elastic and viscous material, resulting to conclude that stress is proportional to a fractional derivative whose order is between zero and one (Fig 5). Hence, from the mathematical point of view, this physical interpretation can be modelled as the following equation.

$$\sigma(t) = X_\alpha D^\alpha \varepsilon(t) \tag{37}$$

Where $0 \leq \alpha \leq 1$ and $X_\alpha$ are the constant coefficients to make an equal relation between stress and strain—the $\alpha$ index represents the dependence of the constant $X$ to expanded order.

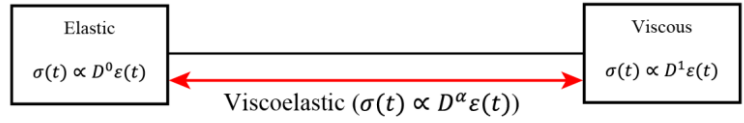

**Fig 5.** A schematic representation of viscoelastic regime and stress-strain relation from a physical point of view.

To obtain the coefficients, the essential condition of fractionalization algorithm (section 2) is applied first. Thus, the equation (41) must contain both Hook's law and Newton fluid in boundary points, i.e. $\alpha = 0$ and $\alpha = 1$. In other words

$$\lim_{\alpha \to 0} X_\alpha = E \quad , \quad \lim_{\alpha \to 1} X_\alpha = \eta \tag{38}$$

One of the specific cases which can satisfy the mentioned boundary condition is $X_\alpha = E^{1-\alpha}\eta^\alpha$. In what follows, it is shown that the proposed $X_\alpha$ is a good suggestion. To make a sufficient condition in fractionalization algorithm, the physical point of view must be considered and to this end, the equation (37) with the assumed coefficients is rewritten.

$$\sigma(t) = E^{1-\alpha}\eta^\alpha \frac{d^\alpha \varepsilon(t)}{dt^\alpha} \tag{39}$$

By applying the Laplace transform to the above FOE the following equation (40) will be obtained—derivative is Caputo.

$$\sigma(s) = E^{1-\alpha}\eta^\alpha s^\alpha \varepsilon(s) \tag{40}$$

As a result, the relaxation modulus will be achieved by considering a unit step strain function.

$$G(t) = \frac{E}{\Gamma(1-\alpha)} \left(\frac{t}{\tau}\right)^{-\alpha}$$

Where $\tau = E/\eta$. Thus, we gained the same equation (40) by the help of physical interpretation and mathematical methods which have better match with the experimental results.

In equation (43), $\alpha$ and $X_\alpha$ are quantities which represent the system property. In other words, it could be possible to consider the $E^{1-\alpha}\eta^\alpha$ coefficient as an independent quantity which is introduced by variable $X_\alpha$—it means that whatever determine the quantity and the structure of the system are just the value of $\alpha$ and $X_\alpha$; indeed, it is possible that a system with $E'$ and $\eta'$ modulus has equal dynamical behavior as another system which has different modulus as $E$ and $\eta$ i.e. $\eta \neq \eta'$ and $E \neq E'$, but $E^{(1-\alpha)}\eta^\alpha = E'^{(1-\alpha)}\eta'^\alpha$. Hence, the product of two moduli which expressed in equation (40) will be just one independent quantity.

If $\alpha \to 0$, the equation (40) will lead to the Hook's law and if $\alpha \to 1$, then the equation turn to Newton fluid model. As a result, based on the physical interpretations, it looks logical to conclude that the proposed model can be a proper tool for investigating the mechanical properties of viscoelastic materials.

To take a more precisely look at the FOE equation in the viscoelastic systems, in what follows, the formalism of inheritance integral for the fractional state is investigated. Equation (41) is obtained by taking partial integral from equation (29).

$$\varepsilon(t) = \sigma(t)J(0^+) + \int_0^t \sigma(t') \frac{dJ(t-t')}{d(t-t')} dt' \tag{41}$$

Also, the creeping state with unit step stress function will be as follow based on FOE equation.

$$J_\alpha(t) = \frac{1}{E} \frac{(t/\tau)^\alpha}{\Gamma(1+\alpha)} \tag{42}$$

By taking derivative from equation (42) and putting the obtained result in equation (41), equation (43) will be achieved.

$$\varepsilon(t) = \sigma(t)J(0^+) + \frac{1}{E\tau^\alpha}\left[\frac{1}{\Gamma(\alpha)}\int_0^t (t-t')^{\alpha-1}\sigma(t')dt'\right] \tag{43}$$

Which the equation inside the bracket is Riemann-Liouville integral; thus, by rewriting the above equation, we have

$$\varepsilon(t) = \sigma(t)J(0^+) + \frac{1}{E\tau^\alpha}\,_0I_t^\alpha \sigma(t) \tag{44}$$

According to the abovementioned subjects, in equation (44), the previous events are embedded in the fractional kernel function $t^{\alpha-1}/\Gamma(\alpha)$; which the obtained result has match with the proposed interpretation of section (3). Regarding those interpretation, the information of a system, which is related to deformation, are storing with a power-law function which has fractional order $\alpha$ from the initial step ($t^+ = 0$) until the present moment ($t$)—which the fractional order is the indication of saving order information. According to the kernel function $t^{\alpha-1}/\Gamma(\alpha)$ in equation (44), $\alpha = 0$ and $\alpha = 1$ are equal with a memory-less system and full-memory system in creeping state respectively owing to $\Gamma(0) = \infty$. Indeed, in a pure viscose system, system's position completely depend on this fact that how stress applies on the system over time. In other words, the nature of stress and the amount of which has a direct impact on the present position of the system. However, owing to this fact that the effect of memory at different times has a unit weight function, the system is able to save all the prior history of itself precisely and equally. In contrast, in the pure elastic system, the present stress is an effective element on the amount of displacement, not previous stresses, because the weight function is zero.

Similarly, the above procedure can be followed for relaxation modulus. The following equation will be obtained by putting equation (36) in (30).

$$\sigma(t) = \varepsilon_0 G(t) + \frac{E\tau^\alpha}{\Gamma(1-\alpha)}\int_0^t (t-t')^{-\alpha}\frac{d\varepsilon(t')}{dt'}dt' \tag{45}$$

As a result, by the help of equation (15) and rewriting the above equation, constitutive equation is gained as follow.

$$\sigma(t) = \varepsilon_0 G(t) + E\tau^\alpha \,_0^C D_t^\alpha \varepsilon(t)$$

Based on the equation (45), in the relaxation state, $\alpha = 0$ and $\alpha = 1$ represent a full-memory system and a memory-less system respectively—$\Gamma(0) = \infty$. In other words, in the pure elastic systems, the displacements of an object in prior times completely keep in the memory of object owing to Because of storing displacement changes over time by an unit function.—known as weight function—over time, and there is no differences in very far and very near time. As a result, all the previous displacements will have been an effect on the value of stress. In the pure viscose system, however, such a memory as the mentioned one does not exist, as well as the previous displacement will not have impacted on the present stress of system; thus, the system is absolutely devoid of any memory in such condition. This is a really intriguing, fascinating interpretation of FOE model with equation (39) because on the one, hand the fractional order $\alpha$ can represent a system which is devoid of any memory, i.e. a memory-less system, when the order is zero, but on the other hand, it can be the representation of a full-memory system.

It is indicated that the fractional models are a valuable tool for describing the dynamic properties of real materials, specifically, in polymers which use for controlling the sounds and vibrations [22-33]. Also, the combination of fractional models with other linear elements can make intriguing structures; for example, it can be referred to fractional Maxwell and Kelvin models which are known as spring-spot models (Table 2). Since SLEs are not always precisely successful in investigating many viscoelastic systems, researchers applied spring-pot models to address the problem, as well as by virtue of these fractional models and the combination of them, they achieved the astonishing results [33-42]—to more details see the "Application" section.

**Table 3.** Fractional order Voigt Models.

| Model | Constitutive equation | Relaxation modulus ($G(t)$) |
|---|---|---|
| 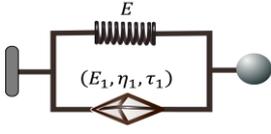 ($E_1, \eta_1, \tau_1$) | $\sigma(t) = E_1 \tau_1^\alpha D^\alpha \varepsilon(t) + E \varepsilon(t)$ | $\dfrac{E_1}{\Gamma(1-\alpha)}\left(\dfrac{t}{\tau_1}\right)^{-\alpha} + E$ |
| 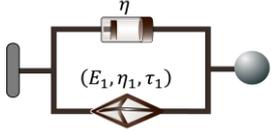 ($E_1, \eta_1, \tau_1$) | $\sigma(t) = E_1 \tau_1^\alpha D^\alpha \varepsilon(t) + \eta D \varepsilon(t)$ | $\dfrac{E_1}{\Gamma(1-\alpha)}\left(\dfrac{t}{\tau_1}\right)^{-\alpha} + +\eta \delta(t)$ |
| 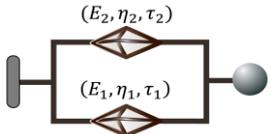 ($E_2, \eta_2, \tau_2$) ($E_1, \eta_1, \tau_1$) | $\sigma(t) = E_1 \tau_1^\alpha D^\alpha \varepsilon(t) + E_2 \tau_2^\beta D^\beta \varepsilon(t)$ | $\dfrac{E_1}{\Gamma(1-\alpha)}\left(\dfrac{t}{\tau_1}\right)^{-\alpha} + \dfrac{E_1}{\Gamma(1-\beta)}\left(\dfrac{t}{\tau_2}\right)^{-\beta}$ |

**Table 4.** Fractional order Maxwell Models

| Model | Constitutive equation | Relaxation modulus ($G(t)$) |
|---|---|---|
| 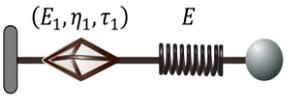 ($E_1, \eta_1, \tau_1$) $E$ | $\sigma(t) + \tau^\alpha D^\alpha \sigma(t) = E \tau^\alpha D^\alpha \varepsilon(t)$ | $E E_\alpha\left[-\left(\dfrac{t}{\tau_1}\right)^{-\alpha}\right]$ |
| 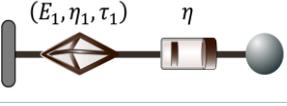 ($E_1, \eta_1, \tau_1$) $\eta$ | $\sigma(t) + \tau^{1-\alpha} D^{1-\alpha} \sigma(t) = \eta_2 D^\alpha \varepsilon(t)$ | $E\left(\dfrac{t}{\tau}\right)^{-\alpha} E_{1-\alpha,1-\alpha}\left[-\left(\dfrac{t}{\tau}\right)^{1-\alpha}\right]$ |
| 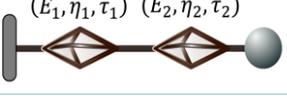 ($E_1, \eta_1, \tau_1$) ($E_2, \eta_2, \tau_2$) | $\sigma(t) + \tau^{\alpha-\beta} D^{\alpha-\beta} \sigma(t) = E \tau^\alpha D^\alpha \varepsilon(t)$ | $E\left(\dfrac{t}{\tau}\right)^{-\beta} E_{\alpha-\beta,1-\beta}\left[-\left(\dfrac{t}{\tau}\right)^{\alpha-\beta}\right]$ |

Assuming the fractional Maxwell model which made of two FOE models with $(\alpha, E_1, \tau_1)$ and $(\beta, E_2, \tau_2)$ elements, where $0 < \alpha, \beta < 1$. Regarding this, the constitutive equation, in relaxation state, will be as follow [to prove, see Appendix]:

$$G(t) = E\left(\dfrac{t}{\tau}\right)^{-\beta} E_{\alpha-\beta,1-\beta}\left[-\left(\dfrac{t}{\tau}\right)^{\alpha-\beta}\right] \quad (46)$$

Where $E_{\alpha,\beta}$ is Mittag-Lefler function [to see the definition of this function, see Appendix]. Above equation will be reduced to the relaxation modulus of classical Maxwell by assuming $\alpha = 0$ and $\beta = 1$. In other words, this equation includes not only the classical relaxation state but also further expansion as compared to the SLEs (more fractional models is presented in the table 3 and 4).

## Periodic strain

As previously detailed, by imposing a periodic strain upon a viscoelastic object, stress is neither exactly in the same phase with strain nor 90º out of phase. In this section, the trueness of this fact is investigated for a FOE model. Assuming a FOE is subjected to a sinusoidal strain, i.e, $\varepsilon = \varepsilon_0 \sin(\omega t)$, then the stress of which will be obtained as follow by the help of equation (39) and Caputo derivative (15).

$$\sigma(t) = E^{1-\alpha}\eta^{\alpha}\left[\frac{\varepsilon_0 \omega}{\Gamma(1-\alpha)}\int_a^t (t-\tau)^{-\alpha}\cos(\omega\tau)d\tau\right] \quad (47)$$

As a result, the following equation hold [to prove the equation, see Appendix].

$$\sigma(t) = \varepsilon_0 E^{1-\alpha}\eta^{\alpha}\omega^{\alpha} i \sin\left(\omega t - \frac{\pi}{2}\alpha\right)\exp\left(i\frac{\pi}{2}(2\alpha-1)\right) \quad (48)$$

In equation (48), if $\alpha = 1$, stress will be 90º out of phase with strain as a pure viscous object, and if $\alpha = 0$, it will be exactly in the same phase with strain as a pure elastic object. Therefore, when $0 < \alpha < 1$, then stress will neither exactly in the same phase with strain nor 90º out of phase. Hence, the result of equation (48) thoroughly matches with physical interpretations.

## Relation between fractional equations and SLE

In this section, with the help of a new fractal system which made of quite a few springs and dampers (Fig 6), the relation between fractional equations and SLEs is investigated.

Increasing the number of elements in the SLEs is one approach to improving the accuracy of a model in the viscoelastic systems. Calculation of and work with such sophisticated systems are not usually easy. Thus, in order to investigate the constitutive equation in infinite systems, which made of many both springs and dampers, different models were derived so far, and their outputs were reduced to fractional equations [24,48,49]. Regarding this, a new fractal system is considered and after massive mathematical calculation, it reveals that the constitutive equation is equal to a spring-pot[11].

Consider the spring-damper circuit in Fig 6, so equation (49) will be obtained with the help of governing equations on the stress/strain elements.

$$\frac{\bar{\sigma}_0^s(s)}{\bar{\varepsilon}_0^d(s)} = \acute{\eta}_0\left(1 - \cfrac{1}{1+\frac{E_0}{\acute{\eta}_1}\left(1+\frac{\acute{\eta}_1}{E_1}\left(1-\cfrac{1}{1+\frac{E_1}{\acute{\eta}_2}\left(1+\frac{\acute{\eta}_2}{E_2}\left(1-\cfrac{1}{\ddots \cfrac{}{1+\frac{E_{n-2}}{\acute{\eta}_{n-1}}\left(1+\frac{\acute{\eta}_{n-1}}{E_{n-1}+\acute{\eta}_n}\right)}}\right)\right)}\right)\right)}\right) \quad (49)$$

Where $\bar{\varepsilon}(s) = L\{\varepsilon(t)\}, \bar{\sigma}(s) = L\{\sigma(t)\}$.

---

[11] From the mathematical point of view, by imposing a local force to the proposed system, an equivalent topology for investigating the relationship between stress and strain over time is found based on the mathematical formulae. In other words, the govern equation in figure 9 is in R space with the Euclidean norm, while the output of the assumed system is located in $L_1$ space.

$$L_1 := \left\{f:[a,b]\to\mathbb{R}; f \text{ is measurable on } [a,b] \text{ and } \int_b^a |f(x)|^1 dx < \infty\right\}$$

Based on equation (50), after rewriting equation (49) to (51), the following equation is obtained[12].

$$x(x+1)^{\alpha-1} = \frac{x}{1+} \frac{(1-\alpha)x}{1+} \frac{\frac{1.(0+\alpha)}{1.2}x}{1+} \frac{\frac{1.(2-\alpha)}{2.3}x}{1+} \frac{\frac{2.(1+\alpha)}{3.4}x}{1+} \frac{\frac{2.(3-\alpha)}{4.5}x}{1+} \ldots \quad (50)$$

$$\frac{\bar{\sigma}_0^s(s)}{\bar{\varepsilon}_0^d(s)} = \acute{\eta}_0 (1 - \frac{c0/s}{1+} \frac{(1-\gamma)c0/s}{1+} \ldots \frac{\frac{(n-1)(n-\gamma)}{(2n-1)(2n-2)}c0/s}{1}) \quad (52)$$

Where $0 < x, \alpha < 1$ and $c_0 = E_0/\eta_0$. The equation (51) will be reduced to the following equation by assuming $c_0 \ll s$ when $n \to \infty$.

$$\frac{\bar{\sigma}_0^s(s)}{\bar{\varepsilon}_0^d(s)} \approx s\eta_0(1 - (c_0/s)^\gamma)$$

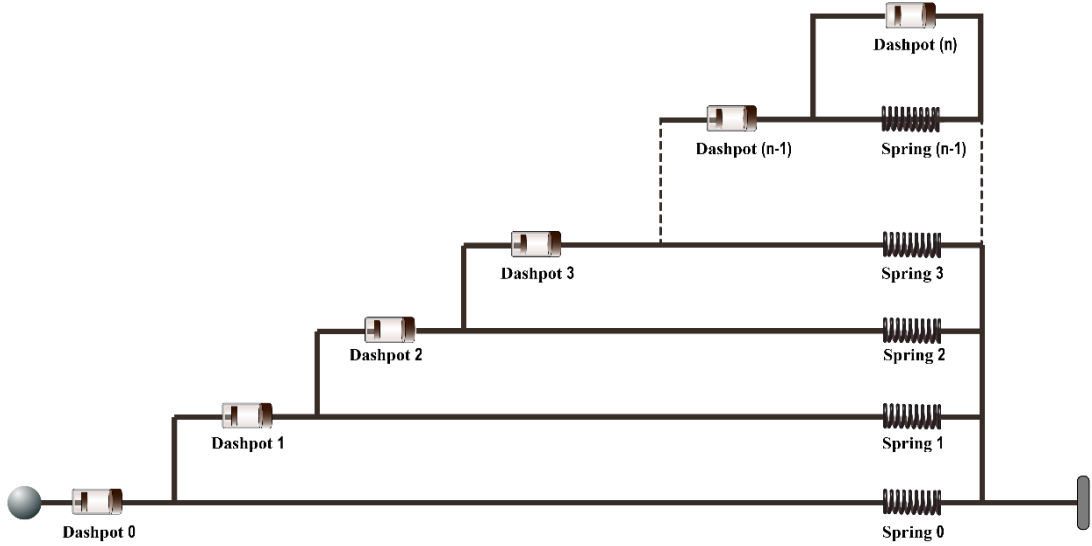

**Fig. 6** Diagram of the finite mechanical arrangement applied to simulate the generalized damper.

By assuming $E_0 = 0$ and rewriting the above equation, we have [to follow the proof in detail, see Appendix]:

$$\sigma^d(t) = 2\eta_0 \frac{d\varepsilon^d(t)}{dt} - \eta_0 \frac{d^{1-2\gamma}\varepsilon^d(t)}{dt^{1-2\gamma}} \quad (54)$$

In other words, the output of the considered system is a combination of a FOE and damper connecting in a series way when $0 < \gamma < \frac{1}{2}$. Also, based on the achieved result, derivative in the above equation is Caputo.

---

[12] Notations

$$\cfrac{a_1}{b_1 + \cfrac{a_2}{b_2 + \cfrac{a_3}{b_3 + \ddots + \cfrac{a_n}{b_n}}}} = \frac{a_1}{b_1+} \frac{a_2}{b_2+} \ldots \ldots \frac{a_n}{b_n}$$

# Application

As the abovementioned sections, fractional calculus is a powerful tool for modeling complex systems, specifically for viscoelastic materials. In this section, at first, a brief review of the application of fractional calculus in the modelisation of viscoelastic materials in the variety fields are presented. In the next step, a one-dimensional fractional model for investigating a cell deformation in creeping state is proposed, and then a 2D fractional model for simulating the network of actin filaments on the cellular surface is proposed. Owing to the perfect match with experimental data and simulated model, we logically conclude that fractional models can be a great replacement for previous models, which are made of spring and damper, in this regard.

Fractional calculus has been employed for modelling viscoelastic systems which cover many fields and subjects. In ref [43], Djordjević, V.D., et al. utilized fractional calculation in four parametric model with equation (53) to investigate a cell's viscoelasticity in the range of $10^{-2} - 10^{2} Hz$ for a periodic test. The output of this model was compared with experimental data obtained from the magnetic oscillatory cytometry method, and it revealed that the proposed model highly matched with experimental data.

$$\sigma(t) = a_0 \varepsilon(t) + a_1 D^\alpha \varepsilon(t) + a_2 \dot{\varepsilon}(t) \tag{53}$$

It is noteworthy Riemann-Liouville (fractional) derivative was considered in equation (53).

Investigation of asphalt mixtures' property during their service life is a real challenge owing to its complexity and sensitivity to environmental and loading conditions. It has been shown that asphalt mixtures behave as linear viscoelastic materials when it is subjected to loading conditions. Traditionally, the viscoelastic material is modeled via creep/recovery functions by the help of spring and damper models. The output of these models have shown an exponential behavior, in nature, however, the mechanical response of these materials have a power-law behavior rather than an exponential. Considering this fact, spring-pot models were proposed for predicting creep/recovery behavior of asphalt mixtures. Finally, the result of proposed models were compared with experimental data, showing that they have in good agreement with the data [39].

Scholars believe that some pulmonary diseases are related to lung viscoelasticity. Thus, profound understanding of viscoelastic models in this field can bring about a new progress in lung pathology and trauma. As a result, stress relaxation test was applied on a pig's lung, and the output of this test showed a power-law behavior than an exponential. Regarding this, two integer standard linear solid— generalized Maxwell models—and a Fractional standard linear solid model were proposed for predicting the mechanical behavior of a system. The result of this investigation indicated that the fractional model by far better matches with experimental data than integer models [41].

To investigate the viscoelastic response of human breast tissue cells, the fractional Zener model was recently proposed [44]. The authors demonstrated that the proposed fractional model has a better result as compared to that of the non-fractional models showed for probing the mechanical response in relaxation state.

According to the recent studies [42], fractional Maxwell and Kelvin model were proposed for analyzing and interpreting the mechanical response of polymer materials, and the result of which have shown that the fractional Maxwell model was not able to investigate the polymer behavior. On the other hand, the fractional Kelvin model is a good proper model with this aim [42].

In 2008, a fractional model was utilized to investigate the arterial viscoelasticity. In that research, the output of the fractional model and SEL model was pinpointed by the help of least-squire. The result that validation showed that the fractional model is able to analyze the system mechanical response to tensile perfectly [35].

## Two-dimension fractional network model

In the last section of the paper, to investigate the mechanical response of a cell, two (new) fractional models are suggested. In the first step, a one-dimension fractional model is proposed, and in the second, a more realistic model is put forward in two dimensions by extending the first model.

We define a cell as the smallest vital unit of the configuration of any live existence. The cell is made of different parts, such as cytoplasm, nucleus, cytoskeleton (CSK), etc. The CSK plays a remarkably essential role in the intracellular force transmission, cellular contractility, as well as the structural integrity of cells not only in the static but in dynamic states as well. The biological functions of cells, such as growth, differentiation, and apoptosis are associated with changes in the cell shape, are related to the mechanical behavior of the CSK [45-47]. The CSK structures are composed of three main elements: actin filaments, intermediate filaments, and microtubules. The structural changes in the CSK, including deformation and rearrangement, are related to changes in the tensile forces that are applied to the actin filaments [45, 46, 48-50]; as a result, the actin filaments have mainly tension-bearing roles.

The viscoelastic properties of cells have been found to play an important role in distinguishing diagnostic, stem cells and etc. Regarding this, several two-dimensional (2D) and three-dimensional (3D) CSK network models have been proposed to calculate the mechanical response of cells [56,58-61].

The viscoelasticity of a cell under some specific external forces was measured with the help of Magnetic Bead Microrheometry methods in creeping state with $F = 1100 pN$ [51]. To investigate the gained results, the author utilized a model which made of two springs and two dampers (Fig 7). The result of this model is compared with the experimental data (Fig 8).

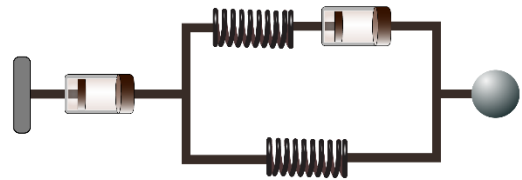

**Fig. 7.** Bausch model has been proposed to investigate a cell viscoelasticity under deformation [51].

The mechanical response of a living cell is quite intricate. The complex, heterogeneous characteristics of cellular structure cause that simple linear models of viscoelasticity cannot predict the mechanical properties of cells under the deformation quite well, especially under minuscule deformation. Regarding this, fractional models are applied to address the point mentioned. In the first step, a FOE model with Eq. 39 is proposed for modeling a cell deformation in the creeping state. To validate the proposed model, the model result is compared with the experimental data which expressed in the ref [51]. In order to make the same condition as experimental data, after solving equation (39) by regarding the creep test, equation (22) is obtained. According to Fig 8, it is clear that the FOE model is a great agreement with the experimental data. By the help of simulated annealing optimization methods, the best coefficients—$\alpha, E, and\ \eta$—in equation (22) is gained as well (Table 5). Hence, it looks logical to conclude that the fractional model can be a suitable replacement for modeling the mechanical properties of a cell than the Bausch model.

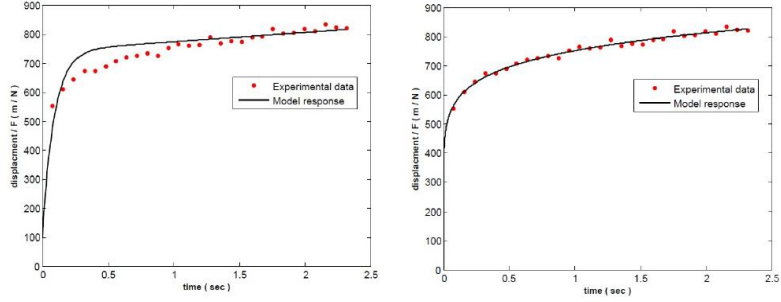

**Fig 8.** Comparison between the FOE model and the Bausch model. The red points and the black line represent the experimental and the response of the FOE models respectively. Which the right side related to the FOE model and the left side represents the Bausch model.

**Table 5**. The optimized coefficients of the FOE model which obtained by annealing optimization methods.

| Fractional order ($\alpha$) | Young's modulus ($Pa$) | Shear modulus ($Pa.s^{-\alpha}$) | MSE |
|---|---|---|---|
| 0.112889249504105 | 0.003456900326295 | 1.207297449819304e-6 | 69.1723 |

To investigate the models' error with experimental data, normalized root-mean-square deviation (NRMSD) method is applied.

$$NRMSD = \frac{\sqrt{MSE}}{y_{max} - y_{max}} \times 100$$

Also, NRMSE was calculated 10.21% and 3.16% for Bausch model and FOE model by the help of above formula respectively, showing the fractional model is highly accurate for probing and predicting the mechanical properties of a subjected cell.

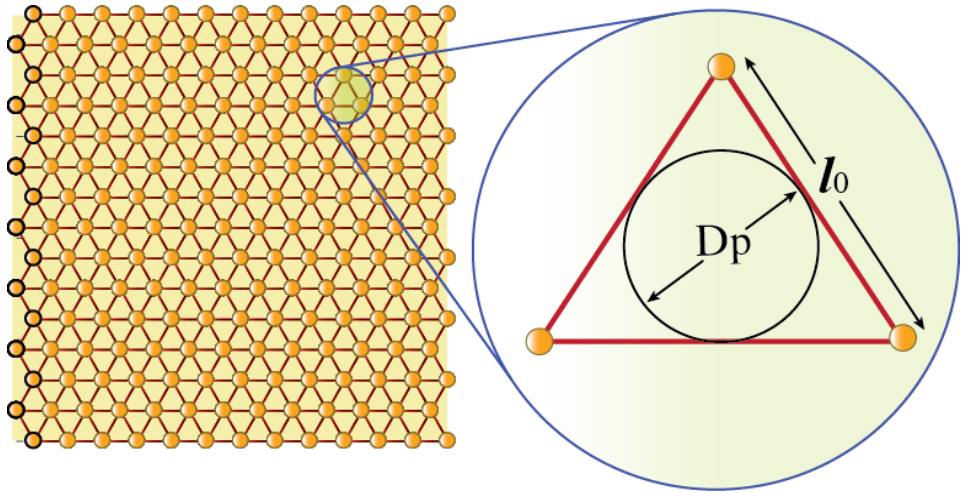

Fig 9. A schematic representation of the preposed network model. Which the black points are fixed and, the others are dynamic. Also, the lines between two points represent the actin filaments connecting by cross-linking protein (in cell cortex).

In the next step, a two-dimension fractional model, so-called 2D fractional network model, is proposed to investigate the cell viscoelasticity. The cell can deal with the external and intercellular forces imposed on it by virtue of CSK. Regarding this, these protein filaments, CSK, is considered as a started point of making a 2D model.

In most cells, actin filaments are intensely concentrated in a layer just beneath the plasma membrane, even though they are found throughout the cytoplasm of a eukaryotic cell. In this region, which is called the cell cortex, these filaments are linked by actin-binding proteins into a meshwork that gives mechanical strength to cells and supports the outer surface of them as well. As a simplification, without loss of generality, imaging the whole mechanical properties of a cell is related to actin filaments [52]. Considering this, we can propose a dynamic network model, shown in Fig. 8, for probing the cell mechanical behavior.

In 2003, a model as the proposed network model in this paper was suggested predicting the mechanical response of a cell [53]. The mechanical properties of actin filaments were approximated with pure elastic materials based on an experimental paper [54]; however, according to the experimental data, these filaments have shown an elastic response in millisecond scale under deformation. Neglecting this scale as a simplification may cause a remarkable error in long time simulation. In other words, the deformation of any material under minuscular stress is divided into three phases which the first one is related to the elastic limit. Also, the mechanical response of an actin filament was reported in this phase, and by overgeneralizing this point, a pure elastic network model—which every actin was approximated with a spring—was proposed in the scale of sec for probing the whole mechanical behavior of a cell [53]. Consequently, the elastic network model, so-called pre-stressed cable network model, was failed to predict the experimental data from bead micrometry method [52]. Regarding this, a new model with FOE elements is suggested in this paper. Two significant merits of which, connected to the system memory and the viscosity term in arbitrary scale time, are embedded in proposed model; resulting the fractional model is able to highly match with the experimental data. The fractional network model is based on the principle that actin filaments, according to ref [52, 55], have shown viscoelastic behavior. Therefore, the FOE model can be a proper model to investigate the mechanical response of each actin filament. By expanding the FOE model into a 2D model with the same order to each element, the fractional network model, which depicts in the Fig. 9, will be obtained.

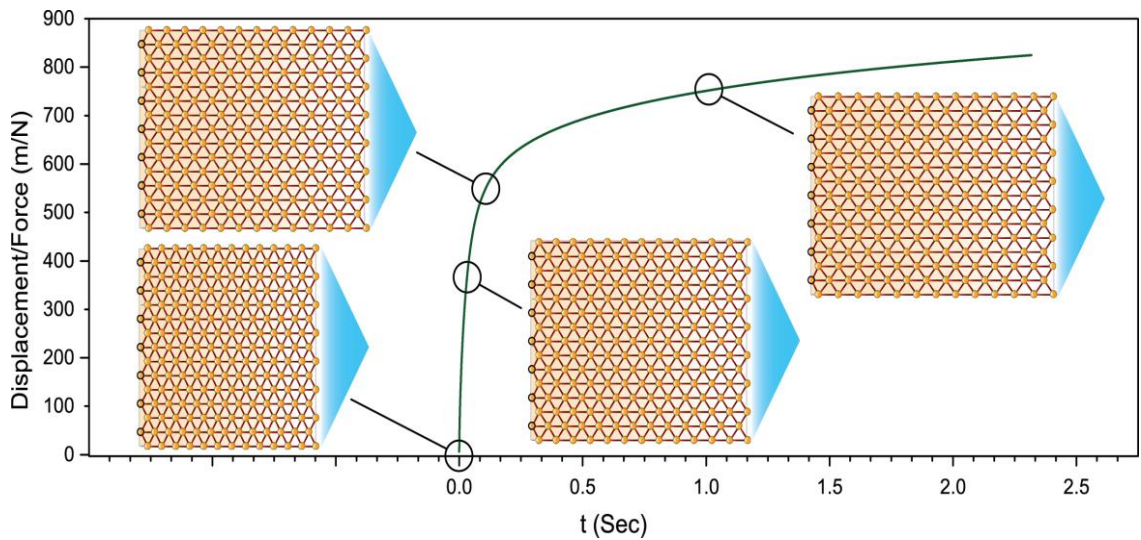

Fig 10. A schematic representation of 2D dynamic fractional simulated network.

Since the direct observation of a $l_0$ (the length of an actin filament) is not exist from a living cell, $l_0$ is approximated from CSK pore ($D_p$) in an adhere cell based on equation $D_p = \sqrt{3}l_0/3$ as paper [53]. In other words, based on the observations, the value of $D_p$ is considered in the middle of experimental data—that is $D_p = 100$nm; resulting $l_0 = 173.2$nm. The space of network is approximated $4.5464 \times 10^6$nm$^2$ based on $l_0$, which is almost same with the reported value in the ref [53].

To validate the simulation result and to compare it with the previous model, we simulate the FOE and spring network, Fig 11 show the behavior of these two model under the same stress. Considering the results, the fractional model is better matched with the experimental data. Hence, it seems that we could conclude that 2D fractional network model is a good candidate for modeling and predicting CSK in cells.

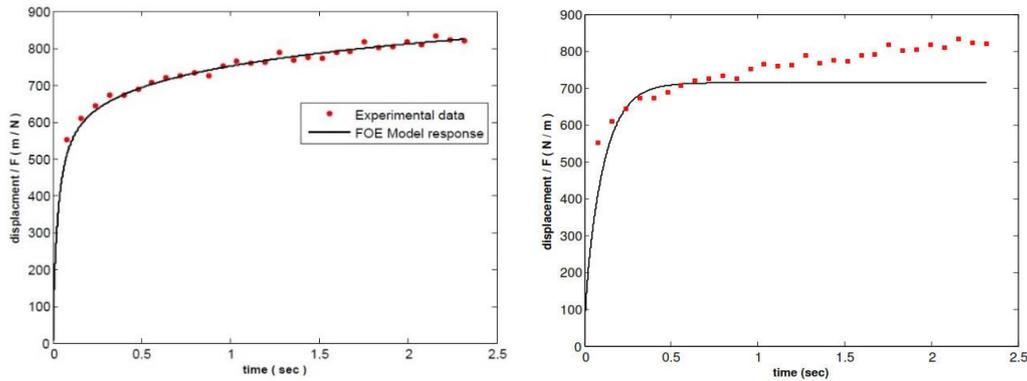

**Fig 11.** A comparison between the 2D dynamic fractional simulated network, the left side figure, and the elastic model which represented in ref [53], the right side figure. Experimental data is obtained from ref [51].

# Discussion

The fractional calculus is a powerful tool for describing the complex physical systems which have long-term memory and long-term spatial interactions. In this paper, different aspects of this powerful tool are introduced through a concise literature review to provide the reader with a picture of what fractional calculus is. Then, with the help of developed concepts, it is sought to find a relation between fractional and viscoelastic systems and to employ the obtained relation for investigating the mechanical property of a cell.

In the first step, to investigate the mechanical response of elastic, viscous, and viscoelastic materials, scholars typically apply spring, damper models and some combination of them respectively. In nature, pure elastic or viscous systems can rarely found and moreover, the mechanical behavior of them are modelled by the help of the Hook's law and Newtonian fluid. Regarding a minuscule deformation, the response of a material is modelled by these laws, thereby making some errors in both the simulation and the calculation of theoretical models—specifically, bio-material such as cells and tissues. In spite of neglecting the arisen errors in some cases, scholars are usually making a highly complex model for improving the accuracy and reducing the error in minute systems. Considering this fact, proposing a new method to cope well with the sophisticated problem can be a worthwhile tool. Based on the investigation which is done in this paper, it is indicated that fractional calculus is a suitable candidate to this end. To prove this claim, a fractal system, made of the combination of quite a few springs and dampers on a large scale, was considered, as well as the output of the proposed system was approximated only with the simple fractional model.

In the next step, a version of fractional models, known as FOE, was employed to model the mechanical response of a cell in one-dimension. By virtue of the achieved result, a 2D model so-called 2D fractional network model was proposed to investigate the mechanical response of actin filaments, exist in CKS and cover a cellular outer layer. The creeping state was considered in both models, and the result of which was validated by not only the experimental date but previous models as well. The results revealed that in the one-dimensional model, data variance relative to a previous model presented in ref [51] was about 11%, while in the FOE model, it was estimated 3%. In other words, the FOE model precision has evaluated 97% in comparison with experimental data. Furthermore, the simulation of two-dimension elastic lattice, represented in ref [53], for actin filaments on the cell surface, is totally invalid in Magnetic Bead Microrheometry method. Meanwhile, the 2D fractional network model has able to analyze same experimental date precisely and to consider the viscosity term of actin filament as well.

The fractional order $\alpha$ in both models is about 0.11. Hence, not only does the investigated model shows more elasticity behavior, but it also has high memory. As a result of done research the paper and many other studies, applying the fractional tool to investigate and to model the mechanical response of phenomena which shows power-law behavior is more efficient than the SLE models.

# Appendix

To facilitate the context for readers, the proof of some equations presenting in the text is presented in this part.

**The proof of Equation 46:**

According to equation (39)

$$\varepsilon_1 = \frac{d^{-\alpha}}{dt^{-\alpha}}\left[\frac{\sigma_1}{E_1 \tau_1^{\alpha}}\right] \quad and \quad \varepsilon_2 = \frac{d^{-\beta}}{dt^{-\beta}}\left[\frac{\sigma_2}{E_2 \tau_2^{\beta}}\right]$$

where the initial value of $\varepsilon_1(t)$ and $\varepsilon_2(t)$ is assumed zero. Considering $\varepsilon = \varepsilon_1 + \varepsilon_2$ based on Maxwell model, we have

$$\frac{1}{E_1 \tau_1^{\alpha}}\frac{d^{-\alpha}\sigma_1}{dt^{-\alpha}} + \frac{1}{E_2 \tau_2^{\beta}}\frac{d^{-\beta}\sigma_2}{dt^{-\beta}} = \varepsilon(t)$$

In addition, $\sigma = \sigma_1 = \sigma_2$ and $0 < \alpha, \beta < 1$ then

$$\sigma_1 + \frac{E_1 \tau_1^{\alpha}}{E_2 \tau_2^{\beta}}\frac{d^{\alpha-\beta}\sigma_2(t)}{dt^{\alpha-\beta}} = E_1 \tau_1^{\alpha}\frac{d^{\alpha}\varepsilon(t)}{dt^{\alpha}}$$

Now let $\tau^{\alpha-\beta} = E_1 \tau_1^{\alpha}/E_2 \tau_2^{\beta}$ and $E = E_1(\tau_1/\tau)^{\alpha}$, so

$$\sigma(t) + \tau^{\alpha-\beta}\frac{d^{\alpha-\beta}\sigma(t)}{dt^{\alpha-\beta}} = E\tau^{\alpha}\frac{d^{\alpha}\varepsilon(t)}{dt^{\alpha}}$$

Taking the Laplace transform and applying the unit step strain function on the above equation, we finally gain the following equation

$$g(s) = \frac{E\tau^{\alpha}s^{\alpha-1}\tau^{\beta-\alpha}}{s^{\alpha-\beta} + (1/\tau)^{\alpha-\beta}}$$

Hence, the following equation by utilizing the inverse Laplace on the above equation is achieved.

$$G(t) = E\left(\frac{t}{\tau}\right)^{-\beta} E_{\alpha-\beta,1-\beta}\left[-\left(\frac{t}{\tau}\right)^{\alpha-\beta}\right]$$

**The proof of Equation 48:**

By taking inverse Laplace from equation (47) and by the help of convolution operation (see table B3.1), the following equation is obtained

$$\sigma(s) = E^{1-\alpha}\eta^{\alpha}\frac{\varepsilon_0\omega}{\Gamma(1-\alpha)}F(s)G(s) \quad s.t \quad F(s) = L\{t^{-\alpha}\}, G(s) = L\{\cos(\omega t)\}$$

according to the table (B3.1)

$$\sigma(s) = \varepsilon_0 \omega E^{1-\alpha}\eta^{\alpha}\frac{s^{\alpha}}{s^2+\omega^2}$$

By taking inverse Laplace from the above equation, we have

$$\sigma(t) = \varepsilon_0 \omega E^{1-\alpha} \eta^\alpha L^{-1}\{\frac{s^\alpha}{s^2+\omega^2}\} \tag{54}$$

In order to solve the above equation, it is enough to calculate $L^{-1}\{\frac{s^\alpha}{s^2+\omega^2}\}$. Thus, according to the (B3.2) equation

$$L^{-1}\{\frac{s^\alpha}{s^2+\omega^2}\} = \frac{1}{2\pi i}\int_{c-i\infty}^{c+i\infty}(\frac{s^\alpha}{s^2+\omega^2})e^{st}ds$$

By assuming $h(z) = z^\alpha e^{zt}/(z^2+\omega^2)$ and Cauchy theorem [56]

$$L^{-1}\{\frac{s^\alpha}{s^2+\omega^2}\} = h(z_1) + h(z_2)$$

Where $z_1 = z_2 = \pm i\omega$ are non-analytical points of function $h(z)$, resulting

$$h(z_1) + h(z_2) = \frac{(i\omega)^\alpha}{2i\omega}e^{i\omega t} + \frac{(-i\omega)^\alpha}{-2i\omega}e^{-i\omega t}$$

$$= \frac{(i\omega)^\alpha}{2i\omega}(e^{i\omega t} - (-1)^\alpha e^{-i\omega t})$$

$$= \frac{\omega^{\alpha-1}}{2}e^{i\frac{\pi}{2}(\alpha-1)}(e^{i\omega t} - e^{i\pi\alpha}e^{-i\omega t})$$

$$= \frac{\omega^{\alpha-1}}{2}(e^{i(\omega t + \frac{\pi}{2}(\alpha-1))} - e^{i(-\omega t + \frac{\pi}{2}(3\alpha-1))}) \tag{55}$$

And also, we have

$$(e^{i(\omega t + \frac{\pi}{2}(\alpha-1))} - e^{i(-\omega t + \frac{\pi}{2}(3\alpha-1))})$$

$$= \cos(\omega t)\cos(\frac{\pi}{2}(\alpha-1)) - \sin(\omega t)\sin(\frac{\pi}{2}(\alpha-1)) - \cos(\omega t)\cos(\frac{\pi}{2}(3\alpha-1)) + \sin(\omega t)\sin(\frac{\pi}{2}(3\alpha-1))$$

$$+ i\left[\sin(\omega t)\cos(\frac{\pi}{2}(\alpha-1)) + \cos(\omega t)\sin(\frac{\pi}{2}(\alpha-1)) - (\cos(\omega t)\sin(\frac{\pi}{2}(3\alpha-1)) - \sin(\omega t)\cos(\frac{\pi}{2}(3\alpha-1)))\right]$$

$$= \cos(\omega t)\left[\cos(\frac{\pi}{2}(\alpha-1)) - \cos(\frac{\pi}{2}(3\alpha-1))\right] - \sin(\omega t)\left[\sin(\frac{\pi}{2}(\alpha-1)) + \sin(\frac{\pi}{2}(3\alpha-1))\right]$$

$$+ i\left[\sin(\omega t)\left[\cos(\frac{\pi}{2}(\alpha-1)) - \cos(\frac{\pi}{2}(3\alpha-1))\right] - \cos(\omega t)\left[\sin(\frac{\pi}{2}(\alpha-1)) + \sin(\frac{\pi}{2}(3\alpha-1))\right]\right]$$

$$= 2\cos(\omega t)\left[\sin(\frac{\pi}{2}(2\alpha-1))\sin\frac{\pi}{2}\alpha\right] - 2\sin(\omega t)\left[\sin(\frac{\pi}{2}(2\alpha-1))\cos\frac{\pi}{2}\alpha\right]$$

$$+ i\left[2\sin(\omega t)\left[\cos(\frac{\pi}{2}(2\alpha-1))\cos\frac{\pi}{2}\alpha\right] - 2\cos(\omega t)\left[\sin\frac{\pi}{2}\alpha\cos(\frac{\pi}{2}(2\alpha-1))\right]\right]$$

$$= 2\sin(\frac{\pi}{2}(2\alpha-1))\left[\cos(\omega t)\sin\frac{\pi}{2}\alpha - \sin(\omega t)\cos\frac{\pi}{2}\alpha\right]$$

$$+ i2\cos(\frac{\pi}{2}(2\alpha-1))\left[\sin(\omega t)\cos\frac{\pi}{2}\alpha - \cos(\omega t)\sin\frac{\pi}{2}\alpha\right]$$

$$= 2\sin(\frac{\pi}{2}(2\alpha-1))\sin(\frac{\pi}{2}\alpha - \omega t) + i\left[2\cos(\frac{\pi}{2}(2\alpha-1))\sin(\omega t - \frac{\pi}{2}\alpha)\right]$$

$$= 2\sin(\omega t - \frac{\pi}{2}\alpha)\left[-\sin(\frac{\pi}{2}(2\alpha-1)) + i\cos(\frac{\pi}{2}(2\alpha-1))\right]$$

$$= 2i\sin(\omega t - \frac{\pi}{2}\alpha)\exp\left(i\frac{\pi}{2}(2\alpha-1)\right) \tag{56}$$

According to (55), (56), and equations (54)

$$\sigma(t) = \varepsilon_0 E^{1-\alpha}\eta^\alpha \omega^\alpha i \, \sin(\omega t - \frac{\pi}{2}\alpha)\exp\left(i\,\frac{\pi}{2}(2\alpha-1)\right)$$

**The proof of Equation 52:**

According to figure (6) the following equations hold.

$$\sigma_k^d = \sigma_{k+1}^d + \sigma_k^s \quad : k = 0,1,\ldots,n-1 \quad (57)$$
$$\varepsilon_k^s = \varepsilon_{k+1}^s + \varepsilon_{k+1}^d \quad : k = 0,1,\ldots,n-2 \quad (58)$$
$$\varepsilon = \varepsilon_0^s + \varepsilon_0^d \quad (59)$$
$$\varepsilon_{n-1}^s = \varepsilon_n^d \quad (60)$$
$$\sigma = \sigma_0^d \quad (61)$$

Regarding Hook's law and Newtonian fluid, we have

$$\sigma^d(t) = \eta \frac{d\varepsilon^d(t)}{dt} \quad (62)$$
$$\sigma^s(t) = E\varepsilon^s(t) \quad (63)$$

where $\sigma^d$, $\varepsilon^d$, $\sigma^s$ and $\varepsilon^s$ related to stress and strain in damper as well as stress and strain in spring respectively. Accordingly, the following equations is obtained by taking Laplace transform from equations (62) and (63).

$$\bar{\sigma}^d(s) = \eta s \bar{\varepsilon}^d(s) \quad (64)$$
$$\bar{\sigma}^s(s) = E\bar{\varepsilon}^s(s) \quad (65)$$

Based on equation (57)

$$\sigma_k^d(t) = \sigma_{k+1}^d(t) + \sigma_k^s(t)$$
$$\overset{(6)}{\to} \eta_k \frac{d\varepsilon_k^d(t)}{dt} = \eta_{k+1} \frac{d\varepsilon_{k+1}^d(t)}{dt} + \sigma_k^s(t)$$
$$\overset{(9)}{\to} \eta_k s \bar{\varepsilon}_k^d(s) = \eta_{k+1} \bar{\varepsilon}_{k+1}^d(s) + \bar{\sigma}_k^s(s)$$
$$\frac{\bar{\sigma}_k^s(s)}{\bar{\varepsilon}_k^d(s)} = \eta_k s - \eta_{k+1} s \frac{\bar{\varepsilon}_{k+1}^d(s)}{\bar{\varepsilon}_k^d(s)} \overset{(9)}{\cong} \eta_k s - \frac{\bar{\sigma}_{k+1}^d(s)}{\left(\frac{1}{\eta_k s}\right)\bar{\sigma}_k^d(s)}$$
$$= \eta_k s - \frac{1}{\left(\frac{1}{\eta_k s}\right) \frac{\bar{\sigma}_k^d(s)}{\bar{\sigma}_{k+1}^d(s)}}$$

Thus, we have

$$\frac{\bar{\sigma}_k^s(s)}{\bar{\varepsilon}_k^d(s)} = \eta_k s \left(1 - \frac{1}{\frac{\bar{\sigma}_k^d(s)}{\bar{\sigma}_{k+1}^d(s)}}\right) \overset{(1)}{\cong} \eta_k s \left(1 - \frac{1}{1+\frac{\bar{\sigma}_k^s(s)}{\bar{\sigma}_{k+1}^d(s)}}\right) \quad (66)$$

In addition, we have

$$\frac{\bar{\sigma}_k^s(s)}{\bar{\sigma}_{k+1}^d(s)} \overset{(10)}{\cong} \frac{E_k \bar{\varepsilon}_k^s(s)}{\bar{\sigma}_{k+1}^d(s)} \overset{(2)}{\cong} E_k \left(\frac{\bar{\varepsilon}_{k+1}^s(s) + \bar{\varepsilon}_{k+1}^d(s)}{\bar{\sigma}_{k+1}^d(s)}\right)$$
$$\overset{(9)}{\cong} \frac{E_k}{\eta_{k+1} s} \left(\frac{\bar{\varepsilon}_{k+1}^s(s) + \bar{\varepsilon}_{k+1}^d(s)}{\bar{\varepsilon}_{k+1}^d(s)}\right)$$

Therefore, the following equation holds

$$\frac{\bar{\sigma}_k^s(s)}{\bar{\sigma}_{k+1}^d(s)} \overset{(10)}{\cong} \frac{E_k}{\eta_{k+1} s} \left(1 + \frac{1}{E_{k+1}} \frac{\bar{\sigma}_{k+1}^s(s)}{\bar{\varepsilon}_{k+1}^d(s)}\right) \quad (67)$$

If we put equation (67) into equation (66), we will obtain

$$\frac{\bar{\sigma}_k^s(s)}{\bar{\varepsilon}_k^d(s)} = \eta_k s \left(1 - \frac{1}{1+\frac{E_k}{\eta_{k+1}s}\left(1+\frac{1}{E_{k+1}}\frac{\bar{\sigma}_{k+1}^s(s)}{\bar{\varepsilon}_{k+1}^d(s)}\right)}\right) \quad (68)$$

In the step $nth$, the following equation will be got

$$\frac{\bar{\sigma}^s_{n-1}(s)}{\bar{\varepsilon}^d_{n-1}(s)} \stackrel{(9)}{=} \eta_{n-1}s\left(\frac{1}{\frac{\bar{\sigma}^d_{n-1}(s)}{\bar{\sigma}^s_{n-1}(s)}}\right) \stackrel{(1)}{=} \eta_{n-1}s\left(\frac{1}{1+\frac{\bar{\sigma}^d_n(s)}{\bar{\sigma}^s_{n-1}(s)}}\right)$$

$$\stackrel{(9,10)}{=} \eta_{n-1}s\left(\frac{1}{1+\frac{\eta_n s}{E_{n-1}}\frac{\bar{\varepsilon}^d_n(s)}{\bar{\varepsilon}^s_{n-1}(s)}}\right) \stackrel{(4)}{=} \frac{\eta_{n-1}s}{\left(1+\frac{\eta_n s}{E_{n-1}}\right)}$$

Thus

$$\frac{\bar{\sigma}^s_{n-1}(s)}{\bar{\varepsilon}^d_{n-1}(s)} = \frac{\eta_{n-1}s}{\left(1+\frac{\eta_n s}{E_{n-1}}\right)} \tag{70}$$

Assuming $\eta_k s = \acute{\eta}_k$ to simplification, so based on equation (68) and (69), in step $(n-1)th$ we will have

$$\frac{\bar{\sigma}^s_{n-2}(s)}{\bar{\varepsilon}^d_{n-2}(s)} = \acute{\eta}_{n-2}\left(1-\frac{1}{1+\frac{E_{n-2}}{\acute{\eta}_{n-1}}\left(1+\frac{\acute{\eta}_{n-1}}{E_{n-1}\left(\frac{\acute{\eta}_n}{E_{n-1}}+1\right)}\right)}\right) \tag{71}$$

By rewriting the above equation

$$\frac{\bar{\sigma}^s_{n-2}(s)}{\bar{\varepsilon}^d_{n-2}(s)} = \acute{\eta}_{n-2}\left(1-\frac{1}{1+\frac{E_{n-2}}{\acute{\eta}_{n-1}}\left(1+\frac{\acute{\eta}_{n-1}}{E_{n-1}+\acute{\eta}_n}\right)}\right) \tag{72}$$

In similar manner, we obtain

$$\frac{\bar{\sigma}^s_{n-3}(s)}{\bar{\varepsilon}^d_{n-3}(s)} = \acute{\eta}_{n-3}\left(1-\frac{1}{1+\frac{E_{n-3}}{\acute{\eta}_{n-2}}\left(1+\frac{\acute{\eta}_{n-2}}{E_{n-2}}\left(1-\frac{1}{1+\frac{E_{n-2}}{\acute{\eta}_{n-1}}\left(1+\frac{\acute{\eta}_{n-1}}{E_{n-1}+\acute{\eta}_n}\right)}\right)\right)}\right) \tag{73}$$

The following equation is achieved in step $(n-4)th$ by the help of above approach

$$\frac{\bar{\sigma}^s_0(s)}{\bar{\varepsilon}^d_0(s)} = \acute{\eta}_0\left(1-\frac{1}{1+\frac{E_0}{\acute{\eta}_1}\left(1+\frac{\acute{\eta}_1}{E_1}\left(1-\frac{1}{1+\frac{E_1}{\acute{\eta}_2}\left(1+\frac{\acute{\eta}_2}{E_2}\left(1-\frac{1}{\ddots \atop 1+\frac{E_{n-2}}{\acute{\eta}_{n-1}}\left(1+\frac{\acute{\eta}_{n-1}}{E_{n-1}+\acute{\eta}_n}\right)}\right)\right)\right)\right)}\right) \tag{74}$$

**NB:** Donating the following symbols to simplification are considered

$$\frac{a_1}{b_1+\frac{a_2}{b_2+\frac{a_3}{b_3+\ddots \atop +\frac{a_n}{b_n}}}} = \frac{a_1}{b_1+}\frac{a_2}{b_2+}\ldots\ldots\ldots\frac{a_n}{b_n} \tag{75}$$

$$\frac{E_{i-1}}{\acute{\eta}_i} = a_i \quad , \quad \frac{\acute{\eta}_i}{E_i} = b_i \tag{76}$$

$$A = \frac{1}{1+a_1\left(1+b_1\left(1-\frac{1}{1+a_2\left(1+b_2\left(1-\frac{1}{\ddots \atop 1+a_{n-1}\left(1+\frac{\acute{\eta}_{n-1}}{E_n+\acute{\eta}_n}\right)}\right)\right)}\right)\right)}$$

Thus

$$A = \frac{1}{[1+a_1(1+b_1)]-}\frac{a_1(1+b_1)}{[1+a_2(1+b_2)]-}\ldots\frac{a_{n-2}(1+b_{n-2})}{1+}\frac{a_{n-1}\left(1+\frac{\acute{\eta}_{n-1}}{E_n+\acute{\eta}_n}\right)}{1} \tag{77}$$

Let us to consider $a_i(1+b_i) = d_i$, by rewriting equation (76), we obtain

$$A = \frac{1}{[1+d_1]-}\frac{d_1}{[1+d_2]-}\ldots\frac{d_{n-2}}{1+}\frac{a_{n-1}\left(1+\frac{\acute{\eta}_{n-1}}{E_n+\acute{\eta}_n}\right)}{1} = \frac{[1+d_1]^{-1}}{1+}\frac{d_1(-[1+d_2])^{-1}}{1+}\ldots\frac{d_{n-2}(1+a_{n-1})^{-1}}{1+}$$
$$\frac{a_{n-1}\acute{\eta}_{n-1}(E_n+\acute{\eta}_n)^{-1}}{1} \tag{78}$$

To take factor form term $1/s$, assuming

$$\acute{a}_i = \frac{E_{i-1}}{\eta_i} \quad , \quad h_i = \frac{E_{i-1}}{E_i} \tag{79}$$

Thus

$$d_i = \frac{1}{s}\acute{a}_i + h_i \quad , \quad a_i = \frac{1}{s}\acute{a}_i \tag{80}$$

Based on equation (76), the equation (44) is rewritten as follow

$$A = \frac{\left[1+h_1+\frac{1}{s}\acute{a}_1\right]^{-1}}{1+} \frac{\left(h_1+\frac{1}{s}\acute{a}_1\right)\left(-\left[1+h_1+\frac{1}{s}\acute{a}_2\right]\right)^{-1}}{1+} \cdots$$

$$\frac{\left(h_{n-2}+\frac{1}{s}\acute{a}_{n-2}\right)\left(1+\frac{1}{s}a_{n-1}\right)^{-1}}{1+} \frac{\acute{a}_{n-1}\eta_{n-1}(E_n+s\eta_n)^{-1}}{1}$$

$$= \frac{\frac{1}{s}[s(1+h_1)+\acute{a}_1]^{-1}}{1+} \frac{\frac{1}{s}(sh_1+\acute{a}_1)\left(-\left[1+h_1+\frac{1}{s}\acute{a}_2\right]\right)^{-1}}{1+} \cdots$$

$$\frac{\frac{1}{s}(sh_{n-2}+\acute{a}_{n-2})\left(1+\frac{1}{s}a_{n-1}\right)^{-1}}{1+} \frac{\frac{1}{s}\acute{a}_{n-1}\eta_{n-1}\left(\frac{1}{s}E_n+\eta_n\right)^{-1}}{1} \tag{81}$$

The following expansion holds according to ref [57].

$$x(x+1)^{\gamma-1} = \frac{x}{1+} \frac{(1-\gamma)x}{1+} \frac{\frac{1\cdot(0+\gamma)x}{1.2}}{1+} \cdots \tag{82}$$

The coefficient of equation (81) is chosen in a way that the equation (82) holds.

$$[s(1+h_1)+\acute{a}_1]^{-1} = c_0 \quad ,$$
$$(sh_1+\acute{a}_1)\left(-\left[1+h_1+\frac{1}{s}\acute{a}_2\right]\right)^{-1} = (1-\gamma)c_0 \quad ,etc \tag{83}$$

Based on equation (63)

$$\frac{\bar{\sigma}_0^s(s)}{\bar{\varepsilon}_0^d(s)} = \acute{\eta}_0\left(1 - \frac{c_0/s}{1+} \frac{(1-\gamma)c_0/s}{1+} \cdots \frac{\frac{(n-1)(n-\gamma)}{(2n-1)(2n-2)}c_0/s}{1}\right)$$

Now let $n \to \infty$ then

$$\frac{\bar{\sigma}_0^s(s)}{\bar{\varepsilon}_0^d(s)} = \acute{\eta}_0\left(1 - \left(\frac{c_0}{s}\right)\left(\frac{c_0}{s}+1\right)^{\gamma-1}\right)$$

Therefore

$$\frac{\bar{\sigma}_0^s(s)}{\bar{\varepsilon}_0^d(s)} \approx s\eta_0\left(1 - \left(\frac{c_0}{s}\right)^\gamma\right) \tag{84}$$

Before applying Laplace transform, the left side of the equation (84) must depend on only a damper. Thus, the following calculation is done to this end.

$$\frac{\bar{\sigma}_0^s(s)}{\bar{\varepsilon}_0^d(s)} \stackrel{(1)}{=} \frac{\bar{\sigma}_0^d(s)}{\bar{\varepsilon}_0^d(s)} - \frac{\bar{\sigma}_1^d(s)}{\bar{\varepsilon}_0^d(s)} \tag{85}$$

Also

$$\frac{\bar{\sigma}_1^d(s)}{\bar{\varepsilon}_0^d(s)} \stackrel{(9)}{=} s\eta_0\left(\frac{1}{\frac{\bar{\sigma}_0^d(s)}{\bar{\sigma}_1^d(s)}}\right) \stackrel{(1)}{=} s\eta_0\left(\frac{1}{1+\frac{\bar{\sigma}_0^s(s)}{\bar{\sigma}_1^d(s)}}\right) \stackrel{(10)}{=} s\eta_0\left(\frac{1}{1+\frac{E_0}{\frac{\bar{\sigma}_1^d(s)}{\bar{\varepsilon}_0^d(s)}}}\right)$$

Considering $\frac{\bar{\sigma}_1^d(s)}{\bar{\varepsilon}_0^d(s)} = y$

$$y = s\eta_0\left(\frac{1}{1+\frac{E_0}{y}}\right)$$

As a result

$$y = s\eta_0 - E_0 \qquad (86)$$

By putting equation (86) into equation (85), we will obtain

$$\frac{\bar{\sigma}_0^d(s)}{\bar{\varepsilon}_0^d(s)} = \frac{\bar{\sigma}_0^s(s)}{\bar{\varepsilon}_0^d(s)} + s\eta_0 - E_0$$

Hence, based on equation (84) we will have

$$\frac{\bar{\sigma}_0^d(s)}{\bar{\varepsilon}_0^d(s)} = s\eta_0\left(1 - \left(\frac{c_0}{s}\right)^\gamma\right) + s\eta_0 - E_0$$

By rewriting the above equation

$$\frac{\bar{\sigma}_0^d(s)}{\bar{\varepsilon}_0^d(s)} = 2s\eta_0 - s\eta_0(c_0/s)^\gamma - E_0 \text{ ;} \qquad (87)$$

By the help of equation (83), we obtain

$$\frac{\bar{\sigma}_0^d(s)}{\bar{\varepsilon}_0^d(s)} = 2s\eta_0 - s\eta_0(s(s(1+h_1) + \acute{a}_1))^{-\gamma} - E_0$$

In other words

$$\frac{\bar{\sigma}^d(s)}{\bar{\varepsilon}^d(s)} = 2s\eta_0 - E_0 - \eta_0 s^{1-\gamma}(s(1+h_1) + \acute{a}_1)^{-\gamma} \qquad (88)$$

By the use of binomial distribution function for every arbitrary $\beta$, the following equation holds.

$$\frac{1}{(1-z)^{\beta+1}} = \sum_{k=0}^{\infty} \binom{k+\beta}{k} z^k$$

Thus

$$(s(1+h_1) + \acute{a}_1)^{-\gamma} = \frac{1}{(s(1+h_1))^\gamma + (1-(-\acute{a}_1/s(1+h_1)))^\gamma} = \frac{1}{(s(1+h_1))^\gamma} \sum_{k=0}^{\infty} \binom{k+\gamma}{k}(-1)^k \left(\frac{\acute{a}_1}{s(1+h_1)}\right)^k \qquad (89)$$

The following equation will have achieved by putting equation (89) into equation (88), Therefor

$$\frac{\bar{\sigma}^d(s)}{\bar{\varepsilon}^d(s)} = 2s\eta_0 - E_0 - \eta_0 s^{1-2\gamma} \frac{1}{(1+h_1)^\gamma} \sum_{k=0}^{\infty} \binom{k+\gamma}{k}(-1)^k \left(\frac{\acute{a}_1}{s(1+h_1)}\right)^k \qquad (90)$$

By the use of Laplace transform from the above equation, we obtain

$$\sigma^d(t) = 2\eta_0 \left(\frac{d\varepsilon^d(t)}{dt}\right) - E_0\varepsilon^d(t) - \eta_0 \frac{1}{(1+h_1)^\gamma} \frac{d^{1-2\gamma}\varepsilon^d(t)}{dt^{1-2\gamma}} +$$

$$\eta_0 \frac{1}{(1+h_1)^\gamma} \sum_{k=0}^{\infty} \binom{k+\gamma}{k}(-1)^k \left(\frac{\acute{a}_1}{1+h_1}\right)^k L^{-1}\{s^{1-2\gamma-k}\bar{\varepsilon}^d(s)\}$$

$$(91)$$

where $0 < \gamma < 1$.

Now let us assume that $E_0 = 0$ in equation (91), resulting in we gain a combination of a damper and FOE connected to each other in a series way.

$$\sigma^d(t) = 2\eta_0 \left(\frac{d\varepsilon^d(t)}{dt}\right) - \eta_0 \frac{d^{1-2\gamma}\varepsilon^d(t)}{dt^{1-2\gamma}}$$

## Acknowledgment

We would also like to show our gratitude to the Dr. Mahdieh Tahmasebi for sharing their pearls of wisdom with us during the course of this research, and we thank Miss Mina Moeini for comments that greatly improved the mathematical proofs in the paper.